\documentclass[12pt]{article}
\usepackage{amsmath}
\usepackage{amsfonts}
\usepackage{ytableau}
\usepackage{cancel}
\usepackage{subfigure}
\usepackage{bm}
\usepackage{tikz}
	\def\be{\begin{equation}}
	\def\bea{\begin{eqnarray}}
	\def\ee{\end{equation}}
	\def\eea{\end{eqnarray}}
\def\d{\partial}

\def\b{\bar}
\def\nn{\nonumber \\}
\def\p{\partial}
\def\t{\tilde}
\def\h{{1\over 2}}
\def\bphi{\b{\phi}}
\def\bpsi{\b{\psi}}
\def \ii {\mathrm{i}}

\def\be{\begin{equation}}
\def\bea{\begin{eqnarray}}
\def\ee{\end{equation}}
\def\eea{\end{eqnarray}}
\def\d{\partial}

\def\b{\bar}
\def\nn{\nonumber \\}
\def\p{\partial}
\def\t{\tilde}
\def\h{{1\over 2}}
\def\bphi{\b{\phi}}
\def\bpsi{\b{\psi}}

\def\be{\begin{equation}}
\def\bea{\begin{eqnarray}}
\def\ee{\end{equation}}
\def\eea{\end{eqnarray}}

\def \a {\alpha}
\def \g {\gamma}

\def \G {\Gamma}
\def \d {\delta}

\def \e {\epsilon}

\def \m {\mu}

\def \t {\tau}
\def \z {\zeta}

\def \cL {{\mathcal L}}

\def \p {\partial}
\def \f {\frac}

\def \hs {\hspace}

\def \lag {\langle}
\def \rag {\rangle}

\def \ii {\mathrm{i}}

\begin{document}
\title{\textbf{Loop Operators in Three-Dimensional $\mathcal{N}=2$ Fishnet Theories}}
\author{
Jun-bao Wu$^{1,2}$,~~
Jia Tian$^2$
and Bin Chen$^{2,3,4}$ \footnote{junbao.wu@tju.edu.cn, wukongjiaozi, bchen01@pku.edu.cn}
}
\date{}

\maketitle
\begin{center}
{\it
$^{1}$Center for Joint Quantum Studies and Department of Physics, School of Science,\\
Tianjin University, 135 Yaguan Road, Tianjin 300350, P.~R.~China\\
$^{2}$Center for High Energy Physics, Peking University, 5 Yiheyuan Rd, Beijing 100871, P.~R.~China\\
$^{3}$Department of Physics and State Key Laboratory of Nuclear Physics and Technology,\\Peking University, 5 Yiheyuan Rd, Beijing 100871, P.~R.~China\\
\vspace{2mm}
$^{4}$Collaborative Innovation Center of Quantum Matter, 5 Yiheyuan Rd, Beijing 100871, P.~R.~China\\
}
\vspace{10mm}
\end{center}

\begin{abstract}
  In this work, we study the line and loop operators in three-dimensional ${\mathcal N}=2$ fishnet theories  in detail. We  construct the straight line and circular loop operators
  which are at least classically half-BPS. We develop a new regularization scheme at frame $-1$ which is suitable  for the study of the fermionic BPS  loops in general super-Chern-Simons-matter theories. We initialize the perturbative computation for the vacuum expectation values of the  circular BPS loop operators   based on this scheme. We construct the cusped line operators  as well, and  compute the vacuum expectation values of these cusped line operators up to two-loop order. We find that the universal cusp anomalous dimension vanishes, if we put aside the fact that the generalized potential has a double pole in the $1/\epsilon$ expansion.
 \end{abstract}
\baselineskip 18pt

\thispagestyle{empty}
\newpage

\section{Introduction}

Wilson loop operators play an important role in the study of dynamics of gauge theory. The vacuum expectation value (VEV) of the Wilson loop
provides  the criteria for color confinement \cite{Wilson:1974sk}. In supersymmetric gauge theory, it is natural to consider the Wilson loop operators
preserving part of the supersymmetries, which have better ultraviolet behavior.  The first kind of  BPS Wilson loop operator was constructed  \cite{Maldacena:1998im, Rey:1998ik}
in four-dimensional  ${\mathcal N}=4$ super Yang-Mills (SYM$_4$) theory. The obtained Maldacena-Wilson
loop has a simple holographic description in terms of classical open string solution \cite{Maldacena:1998im, Rey:1998ik} in the frame work of the AdS/CFT correspondence. This opened a new window to study the AdS/CFT correspondence via the loop operators.

Due to its supersymmetric  nature, the study of the Bogomol'nyi-Prasad-Sommerfield (BPS) Wilson  loop operators benefits from  new techniques in supersymmetric field theories. One typical example is the supersymmetric localization.
 Supersymmetric localization reduces  the path integral computing the VEV of the  circular  half-BPS Wilson loop  in SYM$_4$ to a finite-dimensional integral  \cite{0712.2824}.
 And this finite-dimensional integral turns out be a  Gaussian matrix model. It was known before the localization computations that the results from this Gaussian matrix model are consistent with the prediction of the AdS/CFT correspondence \cite{Erickson:2000af}. Moreover the localization provides a very powerful tool to calculate the VEVs or
certain correlators of the BPS Wilson loops even at finite $N$ (for a collection of reviews, see \cite{1608.02952}).

Besides the smooth line and loop operators, there are the line/loop operators with a cusp, possibly with the insertion of local composite operators. The cusp anomalous dimension is related to  many important quantities  like the ultraviolet divergences of cusped Wilson loops, the infrared divergences of gluon amplitudes and the anomalous dimension of twist-two operators in gauge theories. The study of the cusp anomaly in the BPS line operator could be helped by using the integrability in the $AdS_5/CFT_4$ duality \cite{hep-th/0212208,1012.3982}.  When we insert the composite operators into the Wilson loops, the computation of the anomalous dimensions leads to integrable open spin chains. This happens for both the BPS Maldacena-Wilson loops \cite{hep-th/0604124} and the usual Wilson loops \cite{1810.04643}. The cusp anomalous dimension can be computed based on this open chain firstly using the open asymptotic Bethe ansatz equations and  the boundary thermal Bethe ansatz \cite{1203.1617, 1203.1913} and later using very powerful  tools like the Y-system \cite{1207.5489} and quantum spectral curve  \cite{1510.02098}. The obtained results are consistent with the ones from the localization \cite{1202.4455}. Further studies on comparing results from  localization and  integrability  can be found in \cite{add1, add2}.   Some details of the development on the Wilson loops and integrable structure can be found in  a short review \cite{1911.13141}.

 The supersymmetric localization  has also been applied to the study of the BPS Wilson loop operators in  the AdS$_4$/CFT$_3$ correspondence \cite{Aharony:2008ug},  which states that a three dimensional $\mathcal{N}=6$ Chern-Simons-matter theory is dual to the
IIA string theory on $AdS_4\times CP^3$ background.  In this Chern-Simons-matter theory proposed by Aharony, Bergman, Jafferis and Maldacena (ABJM),  the construction of the BPS Wilson loops is subtle \cite{Drukker:2008zx,Chen:2008bp, Rey:2008bh}. The half BPS one  requires the introduction of superconnection \cite{DT}.
The VEV's of both bosonic $1/6$-BPS \cite{Drukker:2008zx,Chen:2008bp, Rey:2008bh} and fermionic half-BPS Wilson loops \cite{DT} can be computed using the localization  \cite{Hopf} which leads to the results \cite{Marino:2009jd}
 consistent with string theory prediction.  Some fermionic  BPS Wilson loops with less supersymmetries in the ABJM theory were constructed in  \cite{Cardinali:2012ru, Ouyang}. The ABJM theory is also integrable in the planar limit \cite{Minahan:2008, Bak:2008}. There is an interpolating function in the magnon dispersion relation \cite{Gromov:2008qe}.  The computation using the integrability of the cusp anomalous dimension of the Wilson loops was hoped to exactly confirm a conjecture \cite{1403.1894} about  this  interpolating function \cite{1203.1913}.   For a recent progress report on various aspects on the BPS Wilson loops in Chern-Simons-matter theory, see \cite{Road} and \cite{Penati:2020qxp}.

It is certainly interesting to study the  BPS Wilson loops, localizations and integrable structures in the theories with less supersymmetries.
The BPS circular Wilson loops in four dimensional theories need that the theories have at least $\mathcal{N}=2$ supersymmetries. The VEV of  the BPS Wilson loop in the ${\mathcal N}=2^*$ theory, which comes from a mass deformation of the ${\mathcal N}=4$ SYM, was computed using the localizations and the results are consistent with the predictions from the AdS/CFT correspondence \cite{Passerini:2011fe}. This provided an important test of the holographic correspondence in the non-conformal case. 
However, the integrable  $\beta$-deformed SYM$_4$ has only ${\mathcal N}=1$ supersymmetry and it only admits the BPS Wilson line along a null straight line.
The situation is different in three dimensions. The $\beta$-deformed ABJM theory has ${\mathcal N}=2$ supersymmetries so it admits both bosonic half-BPS Wilson loops \cite{Gaiotto:2007qi} and fermionic half-BPS Wilson loops \cite{Ouyang} beyond the lightlike case. Other studies of the BPS Wilson loops in three-dimensional Chern-Simons-matter theories with $2\le {\mathcal N}\le 4$ supersymmetries can be found in \cite{Ouyang:2015qma, Cooke:2015ila,Lietti:2017gtc, Mauri:2017whf,WL}.  It was raised in \cite{Cooke:2015ila} that some classically BPS fermionic Wilson loops may not be truly supersymmetric
at the quantum level. Though there were efforts to resolve this issue either by perturbative calculation \cite{Griguolo:2015swa, FramingABJM, Bianchi:2016vvm} or by studying the precise map between
the BPS Wilson loops and the probe M2-brane solutions in dual M-theory backgrounds \cite{Lietti:2017gtc}, it is fair to say that this big question is still to be answered. It should be
interesting to study this issue in some simple theories with key properties left.

Few years ago, Kazakov et. al. proposed a new double scaling limit on the $\gamma$-twisted $\mathcal{N}=4$ SYM theory
and ABJM theories \cite{Fishnet,FishABJM}. The novel feature in this limit is to take the twisting parameters $q_i$ to
zero (or infinity) and the 't Hooft coupling $\lambda$ to zero but keep the ratio $q_i/\lambda$ (or the product $q_i\lambda$)
of the parameters and the coupling to be finite such that the resulting actions with the gauge field (and gluinos in the $4$D case) being
decoupled include only the  complex scalars and complex fermions. Under this limit, the theories are not unitary and
generically break supersymmetry. However, the resulting new theories  are still integrable in the planar limit,
presenting the remarkable features in the fishnet theories \cite{Zamolodchikov:1980mb,Basso:2017jwq}.
Various aspects \cite{Gromov:2017cja} of these so-called conformal fishnet theories have been studied in the past few years.

In this work, we would like to study the BPS Wilson loop in the three-dimensional (3D) $\mathcal{N}=2$ fishnet theories. A special double-scaling limit of  the $\gamma$-deformed ABJM theory leads to  $\mathcal{N}=2$ fishnet theory with only scalars and fermions \cite{FishABJM}.  We construct the fermionic BPS line and loop operators in this $\mathcal{N}=2$ fishnet theory and compute the VEV of circular loop operators perturbatively after developing a new regularization scheme.
Our computation suggests that one-loop and two-loop contributions to the VEV  are vanishing. Furthermore we investigate the aforementioned line operators with a cusp, which are not supersymmetric anymore.
We find that the  universal cusp anomalous dimension  vanishes at two loop order.

In the perturbative computations of the BPS Wilson loops in the Chern-Simons(-matter) theory,  one important issue  is on the effects of the framing. The perturbative computations  should be performed at framing $-1$ to compare with the prediction from the localization. In this work, our   regularization scheme at framing $-1$  has taken into account of the effects on  the spinors in the definition of the fermionic Wilson loops appropriately. Such improvement could be useful in other settings.


The remaining parts of this paper are organized as follows: in section 2 and 3, we briefly review the three-dimensional $\mathcal{N}=2$ fishnet theories and construct the BPS Wilson loop operators in them respectively. In section 4, we present our detailed perturbative computation on the VEVs of the BPS circular loops operators. In section 5, we discuss the line operator with a cusp. We conclude this paper with some discussions in section 6. We collect some technical details into the appendices.

\section{Three-dimensional $\mathcal{N}=2$ fishnet theories}


Integrable field theories in higher than two dimensions are very rare. In three dimensional  spacetime, the ABJM theory \cite{Aharony:2008ug} was known to be integrable in the planar limit \cite{Minahan:2008, Bak:2008, Gromov:2008qe}.  This theory admits various marginal deformations such as $\beta$-- and $\gamma$--deformations \cite{Beta}, like the four dimensional $\mathcal{N}=4$ SYM theory \cite{Leigh:1995, Lunin:2005}. The $\beta$--/$\gamma$--deformed or twisted theories have less/no supersymmetries while the integrable structures are preserved \cite{Beisert:2005, HeWu, Chen:2016geo}.  Other studies on the spin chain from three dimensional Chern-Simons-matter theories with less or no supersymmetries include the ones in \cite{Bai:2016pxs}.

 In \cite{Fishnet}, a double scaling limit of the $\beta$-- and  $\gamma$--deformed theories was proposed and the resulting theories were called as the fishnet theories due to their fishnet-like Feynman diagrams. One remarkable feature of the fishnet theories is that gauge fields and possible some matter fields in original theories are decoupled so that  the resulted theories are substantially simplified. The fishnet structures of the  Feynman diagrams  indicate  the integrability of these higher dimensional theories \cite{Zamolodchikov:1980mb}.   In this section we will review the  construction of a three-dimensional  fishnet theory with $\mathcal{N}=2$ supersymmetries.

\subsection{ABJM theory and $\gamma$--deformation}
We start with the ABJM theory whose Lagrangian is shown in the Appendix \ref{ABJMaction}. The ABJM theory is a $\mathcal{N}=6$ supersymmetric Chern-Simons-matter theory with gauge group $U(N)\times U(N)$ and the Chern-Simons levels $k$ and $-k$, respectively.  The global symmetry group of this theory is  $OSp(6|4)*U(1)_b$. In particular, the bosonic part of the symmetry group includes the three dimensional conformal group $Sp(4)\sim SO(2,3)$, the R-symmetry group $SU(4)\sim SO(6)$ and an extra $U(1)_b$. The gauge sector consists of two gauge fields $A_\mu$ and $B_\mu$ associated with the first and the second $U(N)$. The matter sector contains the four complex scalars $\phi_I \, (I=1, \cdots, 4)$ and four fermions $\psi^I$ in the bifundamental $(N,\b{N})$ representation.

We perform the $\gamma$--deformation through replacing the ususal product of any two fields $A$ and $B$ with the $\star$ product defined as
\bea
A\star B=e^{\frac{\ii}{2}\mathbf{q}_A\wedge \mathbf{q}_B} AB,
\eea
where the antisymmetric product of the two charge vectors $\mathbf{q}_A$ and $\mathbf{q}_B$ is given by
\bea
\mathbf{q}_A\wedge \mathbf{q}_B=\mathbf{q}_A^T \mathbf{C} \mathbf{q}_B,\quad \mathbf{C}=\left(
\begin{array}{ccc}
	0 & -\gamma _3 & \gamma _2 \\
	\gamma _3 & 0 & -\gamma _1 \\
	-\gamma _2 & \gamma _1 & 0 \\
\end{array}
\right).
\eea
The $U(1)$ charges of the fields are given by the table below:
\bea
\begin{array}{ccccccccc}

	f&\phi_1&\phi_2&\phi_3&\phi_4&{\bpsi_1} &{\bpsi_2}&{\bpsi_3} &{\bpsi_4}\\
	\hline
	q_1& -&+&+&-&-&+&+&-\\
	q_2&+&-&+&-&+&-&+&-\\
	q_3&+&+&-&-&+&+&-&-\\
	\hline
\end{array}
\eea
where $\pm\equiv \pm \h$ (Note that the gauge fields $A_\mu$ and $B_\mu$ are neutral under these three $U(1)$'s.).  This replacement should be performed on every product appearing in the Lagrangian.  The $\gamma$--deformed ABJM action in our convention is presented in the Appendix \ref{ABJMaction}.

\subsection{Double scaling limit and fishnet theory}
Starting from the $\gamma$--deformed ABJM theory,
there are several ways of taking double scaling limits \cite{FishABJM} to get a new theory with only scalars and fermions. The one preserves supersymmetries is the following:
\bea
\gamma_1=\gamma_2=0,\quad e^{-\ii \gamma_3/2}\rightarrow \infty,\quad\lambda\equiv N/k \rightarrow 0,\quad \xi\equiv e^{-\ii \gamma_3}\lambda^2 \text{ fixed}.
\eea
By setting $\gamma_1=\gamma_2=0$ without taking the double scaling limit, the resulting theory is the well known $\beta$--deformed theory with $\mathcal{N}=2$ supersymmetry \cite{Beta}.  After taking the double scaling limit we find that the $\mathcal{N}=2$  supersymmetry is preserved. Under this limit, the gauge sector is decoupled and
the Lagrangian reads
\bea \label{FishLag}
&&\mathcal{L}=\mathcal{L}_k+\mathcal{L}_{Y}+\mathcal{L}_{scalar},
\eea
where the kinetic term is
\bea
\mathcal{L}_k=\text{Tr}(-\p_\mu \bar{\phi}^I \p^\mu \phi_I +\ii\b{\psi}_I \gamma^\mu \p_\mu \psi^I),
\eea
the scalar potential term is
\bea\label{scalar} {\cal L}_{scalar}&=&16\pi^2\frac{\xi}{N^2} \mbox{Tr}(\bphi^1\phi_3\bphi^2\phi_1\bphi^3\phi_2+\bphi^1\phi_3\bphi^4\phi_1\bphi^3\phi_4\nn
&&\bphi^1\phi_2\bphi^4\phi_1\bphi^2\phi_4+{\bphi^2\phi_4\bphi^3\phi_2\bphi^4\phi_3}),
\eea
and the Yukawa-like terms involving the interactions among the scalars and the fermions are
\bea\label{ferm}
 {\mathcal L}_{Y}&=&-2\pi \ii \frac{ \sqrt{\xi}}{N} \mbox{Tr} (-2\bphi^1\phi_3\bpsi_1\psi^3-2\bphi^2\phi_4 \bpsi_2\psi^4\nn
&&-2\bphi^3\phi_2\bpsi_3\psi^2-2 \bphi^4\phi_1\bpsi_4\psi^1
+2 \phi_1\bphi^3\psi^1\bpsi_3+2\phi_2\bphi^4\psi^2\bpsi_4\nn
&&+2\phi_3\bphi^2\psi^3\bpsi_2+2\phi_4\bphi^1\psi^4\bpsi_1+2\bphi^1\psi^4\bphi^2\psi^3-2\bphi^3\psi^1\bphi^4\psi^2\nn
&&-2\phi_1\bpsi_4\phi_2\bpsi_3+2\phi_3\bpsi_1\phi_4\bpsi_2).
\eea
 This fishnet theory is invariant under the $\mathcal{N}=2$ supersymmetries\footnote{We loosely call both Poincar\'e supersymmetry and superconformal symmetry as supersymmetry.} whose transformation rules are summarized as follows,
\bea
&&\delta \phi_1=\ii\b{\epsilon}_{12}\psi^2,~\delta \phi_2=-\ii\b{\epsilon}_{12}\psi^1, \quad \delta \phi_3=\ii\b{\epsilon}_{34}\psi^4,~
\delta \phi_4=-\ii\b{\epsilon}_{34}\psi^3,\nn
&&\delta\bphi^1=\ii\bpsi_2\epsilon^{12},~ \delta\bphi^2=-\ii\bpsi_1\epsilon^{12}, \quad \delta\bphi^3=\ii\bpsi_4\epsilon^{34},~\quad~\delta\bphi^4=-\ii\bpsi_3\epsilon^{34},\nn
&&\delta\psi^1=\gamma^\mu \epsilon^{12}\p_\mu \phi_2+\vartheta^{12} \phi_2+\delta_3\psi^1,\quad \delta\psi^2=-\gamma^\mu \epsilon^{12}\p_\mu \phi_1-\vartheta^{12} \phi_1+\delta_3\psi^2,\nn
&&\delta\psi^3=\gamma^\mu \epsilon^{34}\p_\mu \phi_4+\vartheta^{34} \phi_4+\delta_3\psi^3,\quad \delta\psi^4=-\gamma^\mu \epsilon^{34}\p_\mu \phi_3-\vartheta^{34} \phi_3+\delta_3\psi^4,\nn
&&\delta \bpsi_1=-\b{\epsilon}_{12}\gamma^\mu \p_\mu \bphi^2+\b{\vartheta}_{12}\bphi^2+\delta_3\bpsi_1,\quad\delta \bpsi_2=+\b{\epsilon}_{12}\gamma^\mu \p_\mu \bphi^1-\b{\vartheta}_{12}\bphi^1+\delta_3\bpsi_2, \nn
&&\delta \bpsi_3=-\b{\epsilon}_{34}\gamma^\mu \p_\mu \bphi^4+\b{\vartheta}_{34}\bphi^4+\delta_3\bpsi^3,\quad \delta \bpsi_4=\b{\epsilon}_{34}\gamma^\mu \p_\mu \bphi^3-\b{\vartheta}_{34}\bphi^3+\delta_3\bpsi^4,
\eea
where
\bea\label{delta3}
&& \delta_3\psi^1=-4\pi \frac{\sqrt{\xi}}{N}\epsilon^{34}(\phi_4\bphi^1\phi_3),\quad \delta_3\psi^2=4\pi \frac{\sqrt{\xi}}{N}\epsilon^{34}(\phi_3\bphi^2\phi_4),\nn
&& \delta_3 \psi^3=4\pi \frac{\sqrt{\xi}}{N}\epsilon^{12}(\phi_1\bphi^3\phi_2),\quad \delta_3\psi^4=-4\pi \frac{\sqrt{\xi}}{N}\epsilon^{12}(\phi_2\bphi^4\phi_1),\nn
&& \delta_3 \bpsi_1=4\pi\frac{\sqrt{\xi}}{N} \b{\epsilon}_{34}(\bphi^4 \phi_1 \bphi^3), \quad \delta_3 \bpsi_2=-4\pi\frac{\sqrt{\xi}}{N} \b{\epsilon}_{34}(\bphi^3 \phi_2 \bphi^4),\nn
&& \delta_3 \bpsi_3=-4\pi\frac{\sqrt{\xi}}{N} \b{\epsilon}_{12}(\bphi^1 \phi_3 \bphi^2),\quad \delta_3 \bpsi_4=4\pi\frac{\sqrt{\xi}}{N} \b{\epsilon}_{12}(\bphi^2 \phi_4 \bphi^1).
\eea
The supersymmetry parameters are $\epsilon^{IJ}=\theta^{IJ}+x^\mu\gamma_\mu\vartheta^{IJ}$ and $\b{\epsilon}_{IJ}=\bar{\theta}_{IJ}-\bar{\vartheta}_{IJ}x^\mu \gamma_\mu$, (here $IJ$ only take the values among $12, 21, 34, 43$) with the constraints
\bea
&&\theta^{IJ}=-\theta^{JI}, \quad \bar{\theta}_{IJ}=-\bar{\theta}_{JI}, \quad (\theta^{IJ})^\star=\b{\theta}_{IJ},\quad \bar{\theta}_{12}=\theta^{34},\quad \bar{\theta}_{34}=\theta^{12},\nn
&&\vartheta^{IJ}=-\vartheta^{JI},\quad \b{\vartheta}_{IJ}=-\bar{\vartheta}_{JI}, \quad (\vartheta^{IJ})^\star=\b{\vartheta}_{IJ},\quad \bar{\vartheta}_{12}=\vartheta^{34},\quad \bar{\vartheta}_{34}=\vartheta^{12}.
\eea
In the Euclidean space, the constraints on  the supersymmetry parameters are relaxed to
\bea
&& \theta^{IJ}=-\theta^{JI}, \quad \bar{\theta}_{IJ}=-\bar{\theta}_{JI}, \quad
\bar{\theta}_{12}=\theta^{34},\quad \bar{\theta}_{34}=\theta^{12},\nn
&& \vartheta^{IJ}=-\vartheta^{JI},\quad \b{\vartheta}_{IJ}=-\bar{\vartheta}_{JI}, \quad \bar{\vartheta}_{12}=\vartheta^{34},\quad \bar{\vartheta}_{34}=\vartheta^{12}.
\eea
The Lagrangian of this fishnet theory has been derived in \cite{FishABJM} as a particular example of the plethora of three dimensional fishnet theories. Even though this fishnet theory is not the simplest one which contains the least fields,  the residue supersymmetries provide us another handle to study three dimensional integrable theory. In this note we focus on the supersymmetric (BPS) non-local operators such as the BPS line and loop operators in this fishnet theory.

The above discussions can be also applied to the ABJ theory \cite{ABJ} in which the only difference from the ABJM theory is that the gauge group is now chosen to be $U(N_1)\times U(N_2)$. Here we focus on the case where $N_1$ and $N_2$ are of the same order. Then the $\gamma$-deformation is performed as before. The  fishnet limit preserving $\mathcal{N}=2$ supersymmetries  now becomes,
\bea
&&\gamma_1=\gamma_2=0,\quad e^{-\ii \gamma_3/2}\rightarrow \infty,\quad\lambda_i\equiv N_i/k \rightarrow 0, i=1, 2,\nn
 &&\xi_{(i)}\equiv e^{-\ii \gamma_3}\lambda_i^2\, \hs{2ex}\text{being fixed}, \hs{2ex}i=1, 2.
\eea
The above discussion on the ABJM theory can be carried through straightforwardly with $\frac{\sqrt{\xi}}{N}$ in eqs.~(\ref{ferm}) and (\ref{delta3}) being replaced by $\frac{\xi_{(1)}^{1/4}\xi_{(2)}^{1/4}}{\sqrt{N_1N_2}}$ and $\frac{\xi}{N^2}$ in eq.~(\ref{scalar})  being replaced by $\frac{\sqrt{\xi_{(1)}\xi_{(2)}}}{N_1N_2}$.

\section{BPS Line/Loop Operators}

The  three-dimensional  ${\mathcal N}=2$ fishnet theory obtained from the ABJ(M) theory can be used as a prototype to study various problems in general super-Chern-Simons-matter theories.  We start the investigation with the construction of the BPS line/loop operators in the fishnet theory.  
As in ${\mathcal N}\ge 2$ Chern--Simons--matter theories,  the construction of the fermionic loop operators starts with a superconnection $L$ which combines the bosonic and fermionic field contents together. The supersymmetric (BPS) condition is given by requiring that, under the preserved supercharges, the supersymmetry transformation of  the superconnection can be written as
\bea \label{BPS}\delta L=\p_\tau G+\ii[L,G]\eea for some $G$ \cite{DT, LeeLee}. The BPS conditions are relaxed  comparing  with $\delta L=0$ for the bosonic loop operators. In the following, we still use $W$ to denote the loop operators and reserve $L$ for the superconnection.

 \subsection{Line operators along timelike straight line}

Here and in the following,  we will use the spinor conventions in \cite{WL}. For the Minkowski spacetime with the coordinates $x^\mu=(x^0, x^1, x^2)$ and the metric $\eta_{\mu\nu}=\text{diag}(-1, 1, 1)$,  the $\gamma$ matrices  are chosen to be  $\gamma^{\mu\,\,\beta}_{\,\alpha}=(i\sigma^2, \sigma^1, \sigma^3)$. Following the discussions in \cite{WL}, we can construct  BPS  line operator along the timelike straight line $x^\mu(\tau)=(\tau, 0, 0)$. 
The line operator, which is at least classically half-BPS, is \footnote{In this paper, we only consider taking  trace in the fundamental representation of $U(N_1|N_2)$. }
\be\label{LineOper}
W_{line} ={\textrm{Tr}}( {\mathcal P} \exp(-{\mathrm i} \int d\tau L_{line}(\tau))),
\ee
where \be
L_{line}=\left(\begin{array}{cc}U^I_{~J}\phi_I\bphi^J & \bar{\alpha}_I\psi^I_++\bar{\gamma}_I\psi^I_-\\
\bpsi_{I-}\beta^I-\bpsi_{I+}\delta^I & U^I_{~J} \bphi^J \phi_I\end{array}    \right), \ee
with
\bea && U^I_{~J}=\left(\begin{array}{cccc}\beta^2\bar{\alpha}_2&-\beta^1\bar{\alpha}_2&                          &    \\
                                      -\beta^2\bar{\alpha}_1&\beta^1\bar{\alpha}_1&                          &   \\
                                                            &                     & -\delta^4\bar{\gamma}_4  & \delta^3\bar{\gamma}_4\\
                                                            &                     & \delta^4\bar{\gamma}_3   &  -\delta^3\bar{\gamma}_3   \end{array}\right),\\
&& \bar{\alpha}_I=(\bar{\alpha}_1, \bar{\alpha}_2, 0, 0),~~ \beta^I=(\beta^1, \beta^2, 0, 0), ~~
\bar\gamma_I=(0,0, \bar\gamma_3, \bar\gamma_4), ~~ \delta^I=(0, 0, \delta^3, \delta^4).\eea
 As in \cite{WL}, we define the following bosonic spinors
\be u_{\pm \alpha}=\frac1{\sqrt{2}}\left(\begin{array}{c} 1 \\ \mp {\mathrm i} \end{array}\right), u^\alpha_{\pm}=\frac1{\sqrt{2}} (\mp {\mathrm i}, -1),\label{linespinor}\ee
and decompose a spinor as
\be \label{SpinorDec}\theta_\alpha=u_{+\alpha}\theta_-+u_{-\alpha} \theta_+. \ee

For the above line operators to be BPS, the following constraints
\be\label{BPScondition}\bar\alpha_{1, 2}\delta^{3, 4}=\bar{\gamma}_{3, 4}\beta^{1, 2}=0, \ee
should be satisfied. Then the corresponding preserved supercharges are $\theta^{12}_+, \theta^{34}_-, \vartheta^{12}_+, \vartheta^{34}_-$.
There are two classes of solutions to the BPS constraints in eq.~(\ref{BPScondition})
\bea\label{solution}
\textrm{Class I:}&& \bar{\gamma}^I=\delta_I=0,\\
\textrm{Class II:}&& \bar{\gamma}^I=\beta_I=0,
\eea leading to nontrivial BPS line operators.

\subsection{Circular loop operators}
To study the circular loop operator, we put the theory on the Euclidean space\footnote{To construct BPS circular loop operators, the theory should be put in the Euclidean space \cite{Ouyang:2015ada}. This point was not taken into accout in \cite{Chen:2008bp}.} with the coordinates $x^\mu=(x^1, x^2, x^3)$, the metric $\delta_{\mu\nu}=\text{diag}(1, 1, 1)$
  and the $\gamma$ matrices $\gamma_{\,\alpha}^{\mu\,\,\beta}=(-\sigma^2, \sigma^1, \sigma^3)$. We search for the BPS loop operators along the circle $x^\mu(\tau)=(\cos\tau, \sin\tau, 0)$.

The resulting operator, which is at least classically half-BPS , is,
\be\label{CircularOper}
W_{cir.} = {\textrm{Tr}}({\mathcal P} \exp(-{\mathrm i} \oint d\tau L_{cir.}(\tau))),
\ee
where \be
L_{cir.}=\left(\begin{array}{cc}-{\mathrm i}U^I_{~J}\phi_I\bphi^J & \bar{\alpha}_I\psi^I_++\bar{\gamma}_I\psi^I_-\\
\bpsi_{I+}\delta^I-\bpsi_{I-}\beta^I & -{\mathrm i}U^I_{~J} \bphi^J \phi_I\end{array}    \right), \ee
with the same constants $U^I_{\,\, J}, \bar{\alpha}_I, \beta^I, \bar{\gamma}_I, \delta^I$ and the BPS constraints as  the ones in the previous subsection. Then the corresponding preserved supercharges are the ones satisfying
\be \vartheta_{12}=-{\mathrm i}\gamma_3 \theta_{12}, ~~ \vartheta_{34}={\mathrm i}\gamma_3 \theta_{34}. \ee

Notice that now the definition of $u_\pm$ is changed to
\bea\label{spinorcir} && u_{+\alpha}=\frac{1}{\sqrt{2}}\left(\begin{array}{c} e^{-\frac{{\mathrm i}\tau}{2}}\\ e^{\frac{{\mathrm i}\tau}{2}} \end{array}\right),~~
u_{-\alpha}=\frac{{\mathrm i}}{\sqrt{2}}\left(\begin{array}{c} -e^{-\frac{{\mathrm i}\tau}{2}}\\ e^{\frac{{\mathrm i}\tau}{2}} \end{array}\right), \\
&& u^\alpha_+=\frac{1}{\sqrt{2}}(e^{\frac{{\mathrm i}\tau}{2}}, -e^{-\frac{{\mathrm i}\tau}{2}}), ~~u^\alpha_-=\frac{{\mathrm i}}{\sqrt{2}}(e^{\frac{{\mathrm i}\tau}{2}}, e^{-\frac{{\mathrm i}\tau}{2}}).\eea
And for the circular loop operator, we use the decomposition of the spinor $\theta_\alpha$  \eqref{SpinorDec} with these $u_\pm$.

Based on the discussions in \cite{DT, Ouyang:2015qma, WL}, we can show that classically
\be W_{cir.}-(N_1+N_2)  \ee
is $Q$-exact, where the supercharge $Q$ can be used to perform supersymmetric localization and $N_1,N_2$ are the ranks of the gauge groups. If this relation is preserved at quantum level, we will have
\be <W_{cir.}>=N_1+N_2, \ee
exactly. 

\subsection{${\cal N}=2$ notations}
Later on we will compute the expectation values of the circular and cusped loop/line operators. For the perturbative computations, it is convenient to switch from the ABJM notation to the ${\cal N}=2$ notation \cite{WL},
\be\label{neq2notation}  \phi_I=(Z^1, Z^2, \bar{Z}_3, \bar{Z}_4), \psi^I=(-\zeta^2, \zeta^1, -\bar{\zeta}_4, \bar{\zeta}_3).\ee
As for the $N_{ab}$'s, we have \be N_{12}=N_{21}=2, \, N_{11}=N_{22}=0.  \ee

In  the ${\mathcal N}=2$ notations, the line operators are \eqref{LineOper} with
\bea \label{LineInt}&& L_{line}=B+F, ~~
B= \bar{M}_Z N_{\bar Z}+N_{\bar Z}\bar{M}_Z, ~~
F= \bar{M}_\zeta+N_{\bar\z},\nn
&& [\bar M_Z]_{(ab)} = \bar m^{ab}_i Z_{(ab)}^i, ~~ [N_{\bar Z}]_{(ab)} = n^i_{ab} \bar Z^{(ab)}_{i} ,\nn
&& [\bar M_\z]_{(ab)} = \bar m^{ab}_i \z_{(ab)+}^i, ~~ [N_{\bar\z}]_{(ab)} =  n^i_{ab} \bar\z^{(ab)}_{i-}.
\eea
Note that $\z_{(ab)+}^i = \ii u_+ \z_{(ab)}^i$, $\bar\z^{(ab)}_{i-}= \ii \bar\z^{(ab)}_{i} u_-$ with $u_\pm$ given in eq.~(\ref{linespinor}).

The circular loop operators now in the ${\mathcal N}=2$ notations read \eqref{CircularOper} with
\bea
&& L_{cir.} =   B + F, ~~
   B = - \ii( \bar M_Z N_{\bar Z} + N_{\bar Z} \bar M_Z ), ~~
   F = \bar M_\z - N_{\bar\z}, \nn
&& [\bar M_Z]_{(ab)} = \bar m^{ab}_i Z_{(ab)}^i, ~~ [N_{\bar Z}]_{(ab)} = n^i_{ab} \bar Z^{(ab)}_{i}, \nn
&& [\bar M_\z]_{(ab)} = \bar m^{ab}_i \z_{(ab)+}^i, ~~ [N_{\bar\z}]_{(ab)} =  n^i_{ab} \bar\z^{(ab)}_{i-}.
\eea
Note that $\z_{(ab)+}^i = \ii u_+ \z_{(ab)}^i$, $\bar\z^{(ab)}_{i-}= \ii \bar\z^{(ab)}_{i} u_-$ with $u_\pm$ in eq.~(\ref{spinorcir}).

For both the line and circular loop operators, we have the constraints that \be \bar{m}^{ac}_i\bar{m}^{cb}_j=n_{ac}^in_{cb}^j=0, \ee
with no summations over $c$ in the above equation.
There are two classes of solutions of such constraints,
\bea\label{BPSconditionNeq2}
\textrm{Class I:}&& \bar{m}^{21}_i=n^i_{12}=0,\nn
\textrm{Class II:}&& \bar{m}^{12}_i=n^i_{21}=0,
\eea
which lead to nontrivial line/loop operators.

We can easily switch  between the ABJM notations and the ${\mathcal N}=2$ notations by just finding possibly nonzero components\footnote{The index $i$ in $\bar{m}^{21}_i$ and $n_{12}^i$ only takes the value $3$ or $4$, in order to be consistent with eq.~(\ref{neq2notation}) .} of
$\bar{m}^{ab}_i, n^i_{ab}$ .
For the line operators, they are
\bea && \bar{m}^{12}_1=\bar{\alpha}_2, \hspace{2ex}\bar{m}^{12}_2=-\bar{\alpha}_1, \hspace{2ex}\bar{m}^{21}_3=-\delta^4, \hspace{2ex}\bar{m}^{21}_4=\delta^3,\nn
&& n_{21}^1=\beta^2, \hspace{2ex}n_{21}^2=-\beta^1, \hspace{2ex}n_{12}^3=\bar{\gamma}_4,\hspace{2ex} n_{12}^4=-\bar{\gamma}_3, \eea
And for the circular loop operators, they are
\bea && \bar{m}^{12}_1=\bar{\alpha}_2, \hspace{2ex}\bar{m}^{12}_2=-\bar{\alpha}_1, \hspace{2ex}\bar{m}^{21}_3=\delta^4, \hspace{2ex}\bar{m}^{21}_4=-\delta^3,\nn
&& n_{21}^1=\beta^2, \hspace{2ex}n_{21}^2=-\beta^1, \hspace{2ex}n_{12}^3=-\bar{\gamma}_4, \hspace{2ex}n_{12}^4=\bar{\gamma}_3. \eea
It is then easy to see that the  solutions to the BPS constraints in these two notations map to each other.

\section{Perturbative computations}
In this section we will evaluate the vacuum expectation value (VEV) of the circular  loop operator constructed in the last section up to two-loop level. As we have discussed, the difference between this operator and $N_1+N_2$ are $Q$-exact classically. If this relation is still valid at the quantum level, the vacuum expectation value of this operator will be just $N_1+N_2$, independent of the coupling constants and the parameters in the definition of the loop operators. It will be of great interest to compute this VEV perturbartively  to probe whether the above statement about Q-exactness is valid at the quantum level or not.

Let us first recall some issues arising from  the  perturbative computations of  the BPS Wilson loops in the Chern-Simons-matter theory.
When one tries to compare the perturbative data with the prediction from supersymmetric localization, a subtle issue of framing dependence rises.  The framing dependence of pure Chern-Simons theory has been clearly addressed in \cite{CS, Guadagnini:1989am}. In the perturbation theory,  it comes from the point-splitting regularization and  there it is just a phase factor determined by the linking number of the  Wilson loop contour and the auxiliary contour used for the point-splitting. For the BPS Wilson loops in the super--Chern--Simons theories, the result from the supersymmetric localization is at framing $-1$ \cite{Hopf}. The reason is that the localization is performed by putting the theory on the round $S^3$ and the Wilson loop contour is along a Hopf fiber. The auxiliary  contours used  for the point-splitting should be put on nearby Hopf fibers. After  doing stereographic projection to $\mathbb{R}^3$, the link number between the Wilson loop contour and  any such auxiliary contour is $-1$ \cite{Hopf}. This fact makes the comparison between the results from the localization and the ones from the perturbartive computations complicated since the perturbative calculation is usually done at framing $0$. So we need to modify the perturbative result  by suitably including the effects of framing $-1$ which can be done at lower loop order, or  we need directly perform the perturbative calculations at framing $-1$. In the Chern-Simons-matter theories, the framing dependence is more complicated due to the facts that the framing factors  arising from the non-trivial quantum corrections to the vector propagator  may obtain the contributions from the diagrams with vertices as well \cite{FramingABJM}.  
The perturbative calculations  at such supersymmetric framing  for the fermionic BPS Wilson loops were performed in \cite{WL}  based on the same regularization scheme for the bosonic BPS Wilson loops.

Notice that the definition of the fermionic BPS Wilson loop involves certain constant bosonic spinors. These spinors are the solutions to the Killing spinor equations determined by the Wilson loop contour. When we perform the point-splitting regularization, we move some points from the
Wilson loop to the auxiliary contours as mentioned above. It is reasonable to modify the regularization scheme by  demanding that the spinors corresponding to  these points should also be changed to the solutions of the
Killing spinor equations on these contours. This is the key of our proposal for a new regularization scheme for the computations of the fermionic BPS Wilson loops.

In the following subsection, we will first give the details of our new regularization scheme and later we will use this scheme to compute the VEV
of the circular loop operator in the three-dimensional ${\mathcal N}=2$ fishnet theory.

\subsection{Regularization scheme for Fermionic  loop operators at framing $-1$}


The supersymmetrical localization is first performed on a round $S^3$ \cite{Hopf}.
We parametrize this round $S^3=\{X^i\in \mathbb{R}^4| X^iX_i=1\}$ as
\be X^i=\left(\cos\eta \cos(\tau-\phi), \cos\eta \sin(\tau-\phi), \sin\eta \sin(\tau+\phi), \sin \eta\cos(\tau+\phi)\right). \ee
Here $(\phi, \eta)$ parametrize a $S^2$ and for each fixed pair $(\phi, \eta)$, the $\tau$--circle is the Hopf fiber.
We use the following stereographic projection
\be x^\mu(X^i)=\left(\frac{X^1}{1-X^4}, \frac{X^2}{1-X^4}, \frac{X^3}{1-X^4}  \right), \ee
to map $S^3 \backslash \{(0, 0, 0, 1)\}$ to $\mathbb{R}^3$.
This gives the following parametrization for $\mathbb{R}^3$,
\be x^\mu=\left(\frac{\cos\eta\cos(\tau-\phi)}{1-\sin\eta\cos(\tau+\phi)}, \frac{\cos\eta\sin(\tau-\phi)}{1-\sin\eta\cos(\tau+\phi)}, \frac{\sin\eta\sin(\tau+\phi)}{1-\sin\eta\cos(\tau+\phi)} \right).\ee
Obviously the  $\tau$--circle with $\eta=\phi=0$ gives the Wilson loop contour
\be x^\mu_{WL}(\tau)=(\cos\tau, \sin\tau, 0). \ee
For the auxiliary contour, we can choose $\phi=0, \eta\rightarrow 0$ and keep the terms up to the linear order of $\eta$.
The result is
\be x^\mu_\eta(\tau)=(\cos\tau, \sin\tau, 0)+\eta(\cos^2\tau, \cos\tau\sin\tau, \sin\tau), \ee
which is just the auxiliary contour used in \cite{FramingABJM}. 
The BPS conditions for the spinors $u_{\pm}$ along the auxiliary contour are \cite{WL}
\be \gamma_\mu \dot{x}_\eta^\mu u_{\eta\pm}=\pm |\dot{x}_\eta|u_{\eta\pm}, \quad u_{\eta+}u_{\eta-}=-\ii, \quad u_{\eta\pm} \partial_\tau u_{\eta\mp}=0. \ee
The first equation is a Killing spinor equation along the auxiliary contour.
Demanding that when $\eta\to 0$ these spinors go back to the spinors in eq.~ \eqref{spinorcir}, we get that
\bea && u_\eta(\tau)_{+\alpha}=\frac{1}{\sqrt{2}}\left(\begin{array}{c} e^{-\frac{{\mathrm i}\tau}{2}}(1-\frac{{\mathrm i}}{2}\eta\sin\tau)+\frac\eta2 e^{\frac{{\mathrm i}\tau}2}\\ e^{\frac{{\mathrm i}\tau}{2}}(1+\frac{{\mathrm i}}{2}\eta\sin\tau)-\frac\eta2 e^{-\frac{{\mathrm i}\tau}2} \end{array}\right)+\mathcal{O}(\eta^2),\\
&& u_\eta(\tau)_{-\alpha}=\frac{{\mathrm i}}{\sqrt{2}}\left(\begin{array}{c} -e^{-\frac{{\mathrm i}\tau}{2}}(1-\frac{{\mathrm i}}{2}\eta\sin\tau)+\frac\eta2 e^{\frac{{\mathrm i}\tau}2}\\ e^{\frac{{\mathrm i}\tau}{2}}(1+\frac{{\mathrm i}}{2}\eta\sin\tau)+\frac\eta2 e^{-\frac{{\mathrm i}\tau}2}  \end{array}\right)+\mathcal{O}(\eta^2).\label{spinordeformed}
\eea
These spinors can be also obtained without directly solving the BPS conditions.
Notice that the auxiliary contour $x^\mu_\eta(\tau)$ can be got from the Wilson loop contour $x^\mu_{WL}$ by an affine transformation
\be \tilde{x}^\mu(\tau)=\Lambda^\mu_{\,\,\nu}x^\nu_{WL}(\tau)+a^\mu, \ee
followed by a reparametrization of $\tau=\tau(\tau^\prime)$
\be x^\mu_\eta(\tau^\prime)=\tilde{x}^\mu(\tau(\tau^\prime))+\mathcal{O}(\eta^2),\ee
where \bea \Lambda^\mu_{\,\,\nu}=\left(\begin{array}{ccc}
1 & 0 &0\\
0& \cos\eta& -\sin\eta\\
0&\sin\eta &\cos\eta \end{array}\right), \eea
 is a rotation matrix,
\be a^\mu=(\eta, 0, 0),  \ee
and
\be \tau(\tau^\prime)=\tau^\prime+\eta\sin\tau^\prime+\mathcal{O}(\eta^2). \ee
From  \be u_{\eta}(\tau^\prime)_\pm=\exp(-\frac{\ii}2\eta\gamma_1 )u(\tau(\tau^\prime))_{\pm}, \ee
with $u(\tau)_{\pm}$ being the ones given in eq.~(\ref{spinorcir}) and by keeping only the terms up to the linear order of $\eta$, we find
\bea && u_\eta(\tau^\prime)_{+\alpha}=\frac{1}{\sqrt{2}}\left(\begin{array}{c} e^{-\frac{{\mathrm i}\tau^\prime}{2}}(1-\frac{{\mathrm i}}{2}\eta\sin\tau^\prime)+\frac\eta2 e^{\frac{{\mathrm i}\tau^\prime}2}\\ e^{\frac{{\mathrm i}\tau^\prime}{2}}(1+\frac{{\mathrm i}}{2}\eta\sin\tau^\prime)-\frac\eta2 e^{-\frac{{\mathrm i}\tau^\prime}2} \end{array}\right)+\mathcal{O}(\eta^2),\\
&& u_\eta(\tau^\prime)_{-\alpha}=\frac{{\mathrm i}}{\sqrt{2}}\left(\begin{array}{c} -e^{-\frac{{\mathrm i}\tau^\prime}{2}}(1-\frac{{\mathrm i}}{2}\eta\sin\tau^\prime)+\frac\eta2 e^{\frac{{\mathrm i}\tau^\prime}2}\\ e^{\frac{{\mathrm i}\tau^\prime}{2}}(1+\frac{{\mathrm i}}{2}\eta\sin\tau^\prime)+\frac\eta2 e^{-\frac{{\mathrm i}\tau^\prime}2}  \end{array}\right)+\mathcal{O}(\eta^2),
\eea
as claimed  in
eq.~(\ref{spinordeformed}). This procedure can be performed for a finite $\eta$ with a finite $\phi$ as well, but we would not give the details here.

Now, consider an unregularized integral from the computations of VEV of  the loop operator,
\be \oint d\tau_{1>\cdots>n} I(x(\tau_m), u(\tau_m)), \ee
where $x(\tau_m)=(\cos\tau_m, \sin\tau_m, 0)$ and $u(\tau_m)$ being given in eq.~(\ref{spinorcir}),
and in this section $\oint d\tau_{1>\cdots>n}$ means
\be\int_{2\pi>\tau_1>\cdots>\tau_n>0}\prod_{i=1}^nd\tau_i. \ee To regularize it, we  replace $x(\tau_m), u(\tau_m)$ with
 \be x_m(\tau_m)=(\cos\tau_m+(m-1)\delta\cos^2\tau_m, \sin\tau_m+(m-1)\delta\sin\tau_m\cos\tau_m, (m-1)\delta\sin\tau_m), \ee
and
\bea && u_m(\tau_m)_{+\alpha}=\frac{1}{\sqrt{2}}\left(\begin{array}{c} e^{-\frac{{\mathrm i}\tau_m}{2}}(1-\frac{{\mathrm i}}{2}(m-1)\delta\sin\tau_m)+\frac12(m-1)\delta e^{\frac{{\mathrm i}\tau_m}2}\\ e^{\frac{{\mathrm i}\tau_m}{2}}(1+\frac{{\mathrm i}}{2}(m-1)\delta\sin\tau_m)-\frac12(m-1)\delta e^{-\frac{{\mathrm i}\tau_m}2} \end{array}\right),\\
&& u_m(\tau_m)_{-\alpha}=\frac{{\mathrm i}}{\sqrt{2}}\left(\begin{array}{c} -e^{-\frac{{\mathrm i}\tau_m}{2}}(1-\frac{{\mathrm i}}{2}(m-1)\delta\sin\tau_m)+\frac12(m-1)\delta e^{\frac{{\mathrm i}\tau_m}2}\\ e^{\frac{{\mathrm i}\tau_m}{2}}(1+\frac{{\mathrm i}}{2}(m-1)\delta\sin\tau_m)+\frac12(m-1)\delta e^{-\frac{{\mathrm i}\tau_m}2}  \end{array}\right),
\eea
with $\delta$ the regularization parameter\footnote{it is just the $\eta$ in the previous discussions.}.
This is our proposal of the regularization scheme for the  circular BPS   loop operators at framing $-1$.

\subsection{The vacuum expectation values of circular loop operators}

In the perturbative computations, we assume\footnote{Recall that we always assume that $N_1$ and $N_2$ are at the same order.} $\bar{m}_i^{ab}$ and $n^i_{(ab)}$ to be of  order ${\cal O}(\xi^{1/4}/\sqrt{N_1})$ .  We make this choice partly because  this is similar to what happens for the half-BPS Wilson loops in the ABJM theory. Then up to two-loop order, there are no contributions from the diagrams with vertices. All the relevant diagrams have appeared in the previous computations \cite{Drukker:2008zx, Chen:2008bp, Rey:2008bh, Bianchi:2013zda, WL}, but here for the diagrams with fermionic propagators we would apply  our new regularization scheme.
\begin{figure}[htbp]
  \centering
  \includegraphics[height=0.16\textwidth]{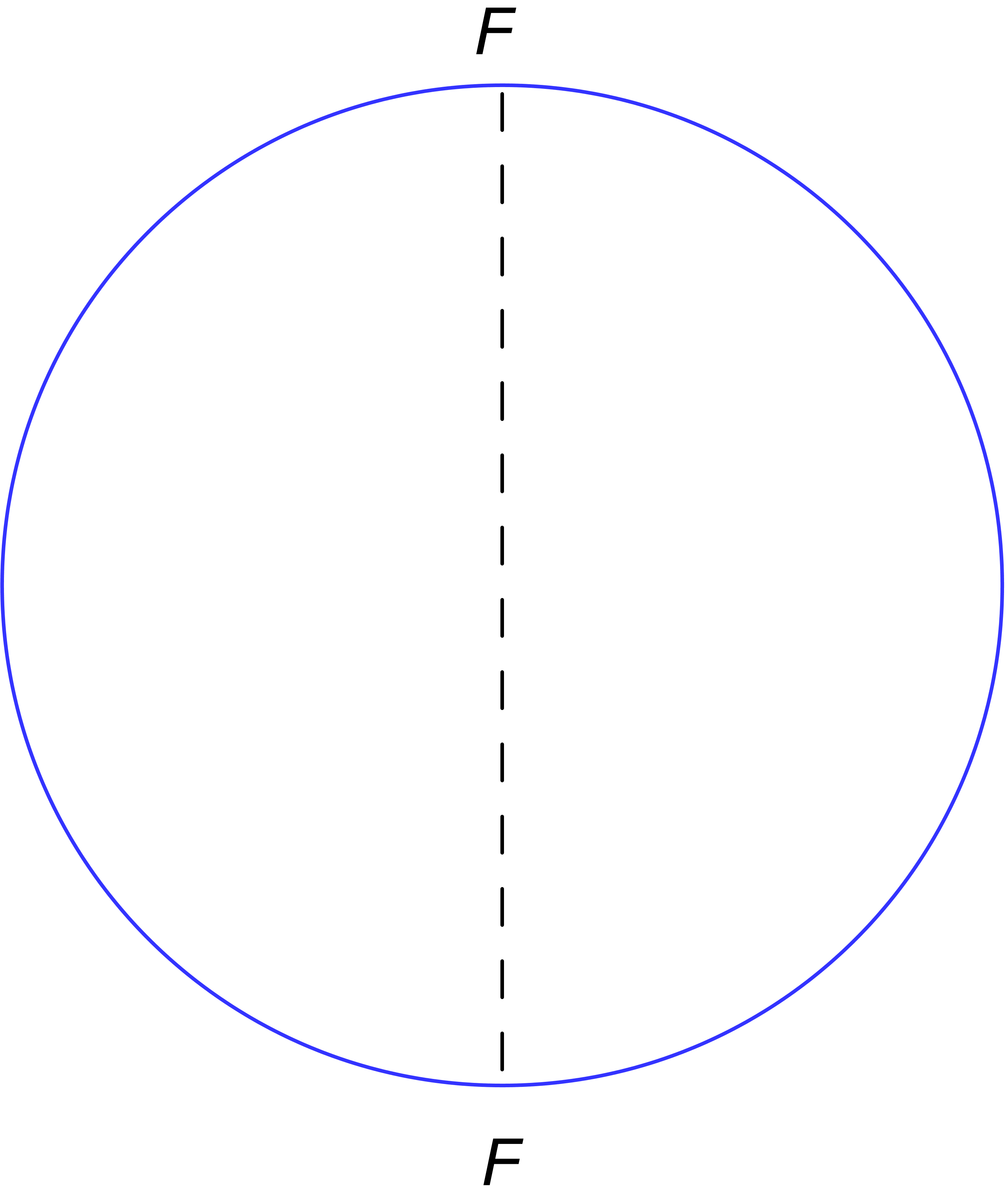}
  \caption{One-loop Feynman diagram.} \label{fdFF}
\end{figure}
The only contributing  one-loop diagram is from exchanging  one fermions (fig.~\ref{fdFF}),
\bea   W_1&=&\sum_{a, b}\bar{m}^{ab}_i n_{ba}^i N_a N_b {\mathcal I}_1^{(-1)},\nn
&=&\sum_{a\ne b}\bar{m}^{ab}_i n_{ba}^i {\mathcal I}_1^{(-1)}.\eea
with the integral being\footnote{The factor $|\dot{x}_i(\tau_i)|$ is introduced to make the integral be reparametrization invariant of $\tau_i$. This could be understood as part of the regularization scheme at framing $-1$. They do not appear in the construction of the loop operators since we have chosen $\tau$ such that we always have $|\dot{x}_i|=1$ there.}

\bea
 {\mathcal I}^{(-1)}_1= - \ii \f{\G(\f32-\e)}{2\pi^{\f32-\e}}
                              \!\oint\! d\t_{1>2}
                              \Big( \f{|\dot x_1(\tau_1)||\dot x_2(\tau_2)|u_1(\t_1)_+\g_\m u_2(\t_2)_-(x^\mu_1(\tau_1)-x^\mu_2(\tau_2))}{|x_1(\tau_1)-x_2(\tau_2)|^{3-2\e}} - (1\leftrightarrow 2) \Big).
\eea

There are numerical evidences that the results for the   integral in the above equation  is the same as the one from the dimensional regularization with dimensional reduction   at framing $0$  when $\epsilon>1/2$. We take the continuation of the later to $\epsilon=0$ as the results in our new supersymmetic regularization, then  we are led to the claim that
\be  {\mathcal I}^{(-1)}_1=0.\ee
\begin{figure}[htbp]
  \centering
  \subfigure[]{\includegraphics[height=0.16\textwidth]{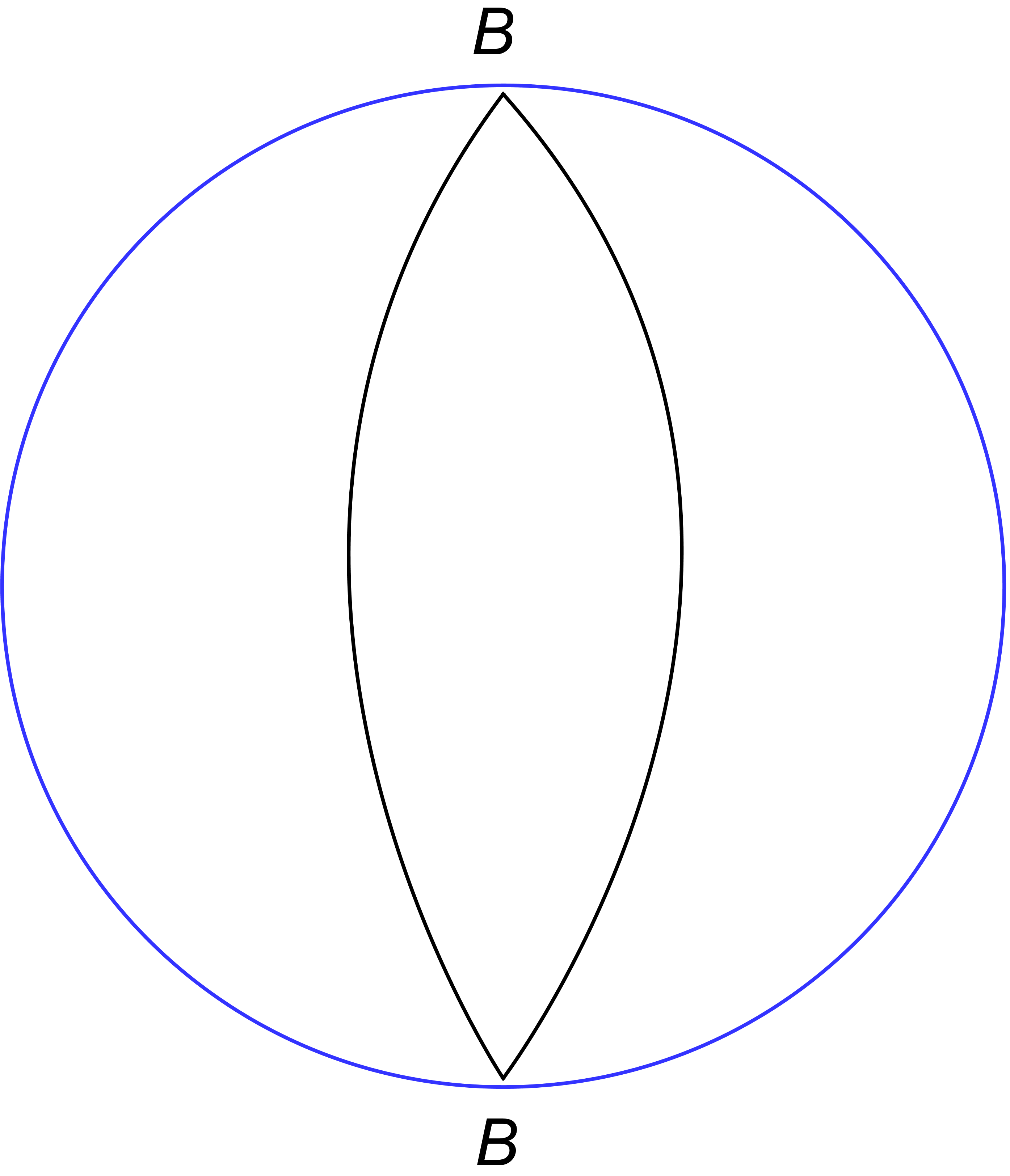}\label{fdBB}}\quad
  \subfigure[]{\includegraphics[height=0.16\textwidth]{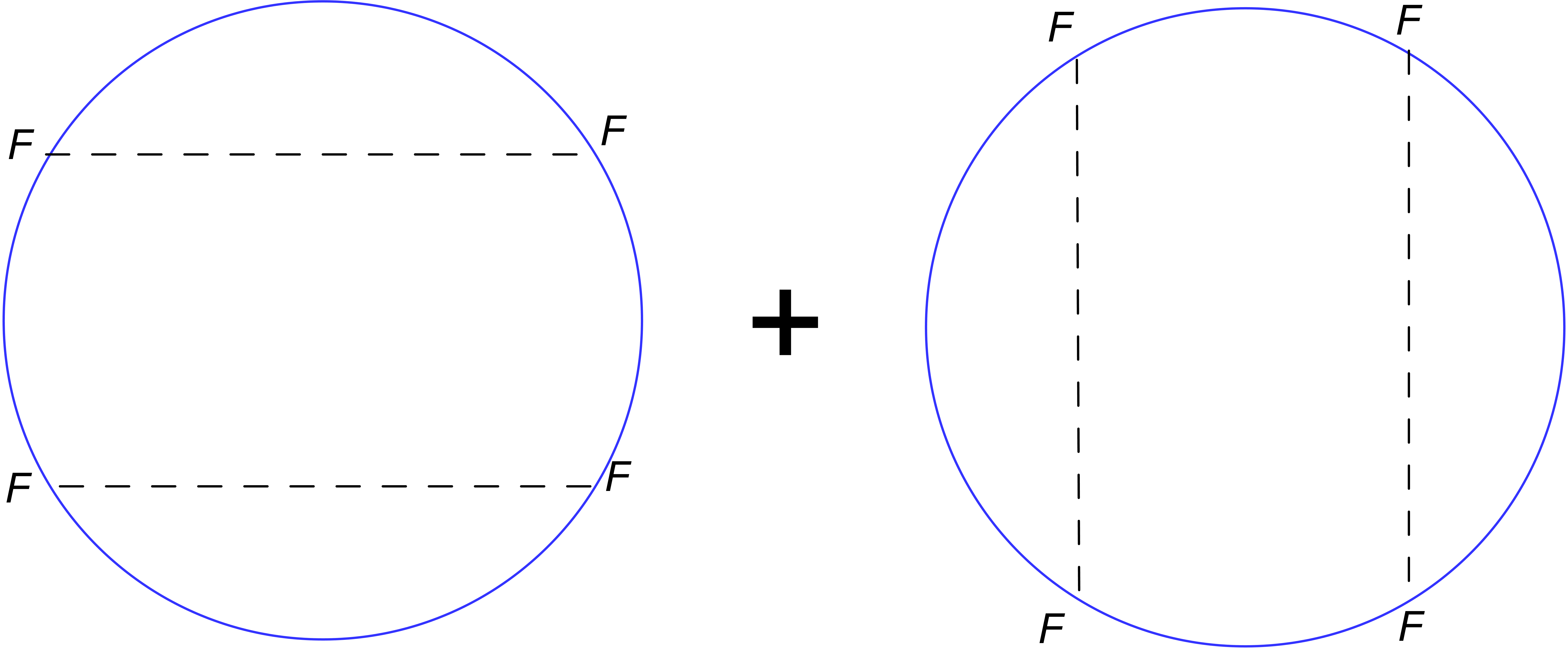}\label{fdFFFF}}\\
  \caption{Two-loop Feynman diagrams.}\label{fd5}
\end{figure}
One part of the two-loop  contributions comes from the diagrams with two scalar propagators (fig.~\ref{fdBB}), which give
\bea W_{2; 1} &=& \sum_{a, b, c}(\bar{m}^{ab}_in^i_{ba}\bar{m}^{ac}_jn^j_{ca}+\bar{m}^{ba}_in^i_{ab}\bar{m}^{ca}_jn^j_{ac})N_aN_bN_c{\mathcal I}^{(-1)}_{2a}\\
&=&N_1N_2(N_1+N_2)\sum_{a\ne b}(\bar{m}_i^{ab}n^i_{ba})^2 {\mathcal I}^{(-1)}_{2a}.\eea
with
\be {\mathcal I}^{(-1)}_{2a} =  \f{\G^2(\f12-\e)}{16\pi^{3-2\e}}
                                \!\oint\! d \t_{1>2} \f{|\dot x_1(\tau_1)| |\dot x_2(\tau_2)|}{|x_1(\tau_1)-x_2(\tau_2)|^{2-4\e}}.  \ee
It can be shown that it  vanishes at any framing, not only at framing $-1$.

The other part coming from the diagrams with two fermion propagators (fig.~\ref{fdFFFF})  gives
\bea W_{2; 2}&=&\sum_{a, b, c}N_aN_bN_c (\bar{m}^{ab}_in^i_{ba}\bar{m}^{ac}_jn^j_{ca}{\mathcal I}^{(-1)}_{2b}+\bar{m}^{ba}_in^i_{ab}\bar{m}^{ca}_jn^j_{ac}{\mathcal J}^{(-1)}_{2b})\\
&&=N_1N_2(N_1+N_2)\sum_{a\ne b}(\bar{m}_i^{ab}n^i_{ba})^2({\mathcal I}^{(-1)}_{2b}+{\mathcal J}^{(-1)}_{2b}), \eea
where
\bea && {\mathcal I}^{(-1)}_{2b} = - \f{\G^2(\f32-\e)}{4\pi^{3-2\e}}
                                   \!\oint\! d\t_{1>2>3>4}
                                   \Big(
                                     \f{|\dot x_1(\tau_1)||\dot x_2(\tau_2)|
                                        u_1(\t_1)_+\g_\m u_2(\t_2)_-(x_1^\mu(\tau_1)-x_2^\mu(\tau_2))}
                                       {|x_1(\tau_1)-x_2(\tau_2)|^{3-2\e}} \nn
&& \phantom{{\mathcal I}^{(-1)}_{2b} =}
                                     \times
                                     \f{|\dot x_3(\tau_3)||\dot x_4(\tau_4)|
                                        u_3(\t_3)_+\g_\m u_4(\t_4)_-(x_3^\mu(\tau_3)-x_4^\mu(\tau_4))}
                                       {|x_3(\tau_3)-x_4(\tau_4)|^{3-2\e}}
                                     - (1234\rightarrow 2341)
                                    \Big),
\eea

\bea && {\mathcal J}^{(-1)}_{2b} = - \f{\G^2(\f32-\e)}{4\pi^{3-2\e}}
                                   \!\oint\! d\t_{1>2>3>4}
                                   \Big( \f{|\dot x_1(\tau_1)||\dot x_2(\tau_2)|
                                        u_2(\t_2)_+\g_\m u_1(\t_1)_-(x_2^\mu(\tau_2)-x_1^\mu(\tau_1))}
                                       {|x_1(\tau_1)-x_2(\tau_2)|^{3-2\e}}
                                    \nn
&& \phantom{{\mathcal J}^{(-1)}_{2b} =}
                                     \times
                                     \f{|\dot x_3(\tau_3)||\dot x_4(\tau_4)|
                                        u_4(\t_4)_+\g_\m u_3(\t_3)_-(x_3^\mu(\tau_3)-x_4^\mu(\tau_4))}
                                       {|x_3(\tau_3)-x_4(\tau_4)|^{3-2\e}}
                                     - (1234\rightarrow 2341)
                                    \Big).
\eea
It is easy to get that \be
{\mathcal I}^{(-1)}_{2b}={\mathcal J}^{(-1)}_{2b}.\ee
We expect them  to be zero, but we  have only weak numerical evidence for this.
If it is the case, we have \be \langle W \rangle =N_1+N_2,\ee
up to two-loop order.

\section{Line operator with a cusp}
The cusp anomalous dimension  plays an important role in gauge theories. It is related to many important quantities like the ultraviolet divergences of the cusped Wilson loops, the infrared divergences of the gluon amplitudes and the anomalous dimension of the twist-two operators. The generalized Wilson loops in ABJ(M) have been constructed  in \cite{GenCusp}  and their VEV's have been  computed there up to two loops.\footnote{Studies on related Bremsstrahlung functions can be found in \cite{1705.10780}. } Here we construct the cusped line operators in three-dimensional ${\cal N}=2$ fishnet theory and compute their VEV's.

\subsection{The construction}

After doing Wick rotation to the Euclidean space, the BPS line in  the ${\mathcal N}=2$ notations becomes \eqref{LineOper} and \eqref{LineInt} with the replacement
\bea
L_{line}=\ii B+F.
\eea
We consider a cusp which is parametrized by
\bea\label{cusp}
x^1=\tau \cos\phi,\quad x^2=|\tau| \sin\phi,\quad x^3=0,\quad -L  \leq \tau \leq L,\quad 0\leq \phi<\frac{\pi}{2}.
\eea
 The constant spinors in the definition of the cusped line operator is given by applying the proper rotation
\bea\label{cuspspinor}
S_{R}=e^{-\phi \gamma_2/2}=\left(
\begin{array}{cc}
	e^{-\frac{\ii \phi }{2}}  & 0 \\
	0 & e^{\frac{\ii \phi }{2}} \\
\end{array}
\right),\quad S_{L}=e^{\phi \gamma_2/2}=\left(
\begin{array}{cc}
	e^{\frac{\ii \phi }{2}}  & 0 \\
	0 & e^{-\frac{\ii \phi }{2}} \\
\end{array}
\right)
\eea
to the constant spinors $u_{\pm \alpha}$ in eq.~(\ref{linespinor}).
The results are
\bea
&&\mbox{Right-half}:~u_{+\alpha, R}=\frac1{\sqrt{2}}\left(\begin{array}{c} s_-\\ -\ii s_+\end{array}\right),\quad u_{-\alpha, R}=\frac1{\sqrt{2}}\left(\begin{array}{c} s_-\\ \ii s_+ \end{array}\right),\\
&&\mbox{Left-half}:~u_{+\alpha,L}=\frac1{\sqrt{2}}\left(\begin{array}{c} s_+\\ -\ii s_-\end{array}\right),\quad u_{-\alpha, L}=\frac1{\sqrt{2}}\left(\begin{array}{c} s_+\\ \ii s_- \end{array}\right),
\eea
where we have defined $s_{\pm}=\exp(\pm \ii\phi/2)$.
It is easy to check that along both the left and the right half the following BPS conditions are satisfied
\be \gamma_\mu \dot{x}^\mu u_{\pm}=\pm |\dot{x}|u_{\pm}, \quad u_{+}u_{-}=-\ii, \quad u_{\pm} \partial_\tau u_{\mp}=0. \ee
Finally, the cusp  operators are defined by taking the contour in eq.~(\ref{cusp}) and the spinors in eq.~(\ref{cuspspinor}). We also need to keep in mind the BPS constraints in eq.~(\ref{BPSconditionNeq2}) whose solutions $\bar{m}_i^{21}=n^i_{12}=0$ and $\bar{m}_i^{12}=n^i_{21}=0$ lead to nontrivial cusp operators.
Since the left and right part of the cusp operators preserves different supersymmetry the whole cusp configuraton is not BPS. One may try to search for  the BPS generalized cusp  as in \cite{GenCusp} in the fishnet theory. But we find that a direct generalization of the construction introduced in \cite{GenCusp} does not  lead to the BPS configuration. This is reasonable   considering the fact that the fishnet theory has less supersymmetries than the ABJM theory. 

\subsection{Perturbative calculations}
In this subsection, we compute the VEV's of  the cusp operators constructed in the previous subsection. For simplicity, we do the computations in the framing $0$ using the dimensional regularization with dimensional reduction and take the internal data $\bar{m}_i^{ab}, n_{ab}^i$ being the same along the cusp. As in the circular loop case, we assume $\bar{m}_i^{ab}$ and $n^i_{ab}$ to be  of order $\mathcal{O}(\xi^{1/4}/\sqrt{N})$. Almost all of the needed integrals have already been computed in \cite{GenCusp}, so we skip the details of the calculations for these integrals. Notices that we should keep the finite piece in the one-loop result.
\begin{figure}[htbp]
  \centering
  \subfigure[]{\includegraphics[height=0.16\textwidth]{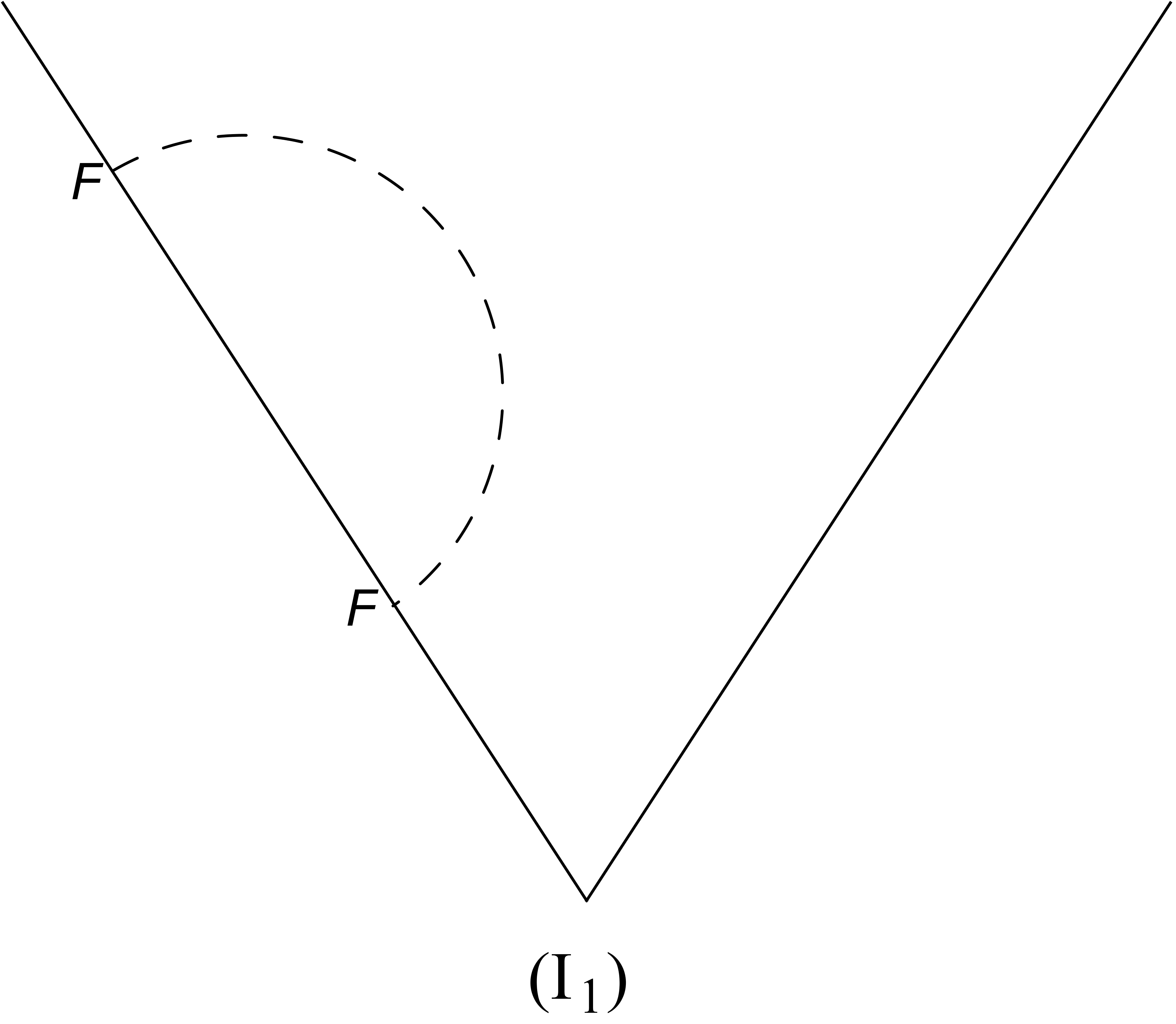}\label{I1}}\qquad \qquad
  \subfigure[]{\includegraphics[height=0.16\textwidth]{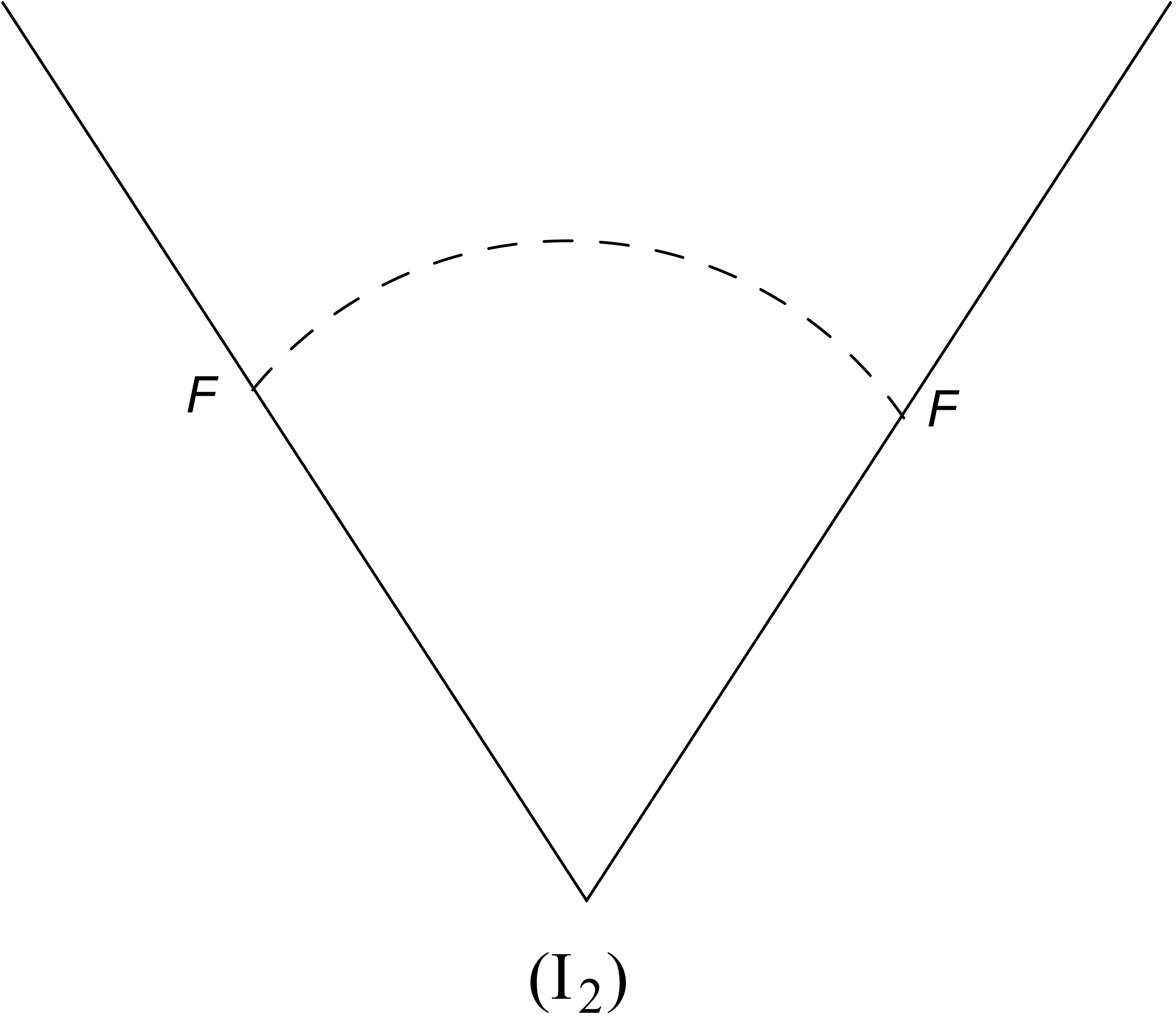}\label{I2}}\\
  \caption{One-loop Feynman diagrams for cusp.}
\end{figure}
The one-loop contributions come from two kinds of diagrams with a single fermion propagator ending on the same line (fig.~\ref{I1}) or the different lines (fig.~\ref{I2}). 
The sum of their contributions is
\bea
W_1
=\sum_{a, b}\bar{m}^{ab}_i n^i_{ba}N_aN_b(2~{\mathcal I}_1+{\mathcal I}_2),
\eea
with
\bea
\mathcal{I}_1
&=&-\frac{\Gamma(\frac{3}{2}-\epsilon)}{\pi^{\frac{3}{2}-\epsilon}} \int_{-L<\tau_2<\tau_1<0} \frac{d\tau_1d\tau_2}{(\tau_1-\tau_2)^{2-2\epsilon}}=\frac{1}{4\epsilon}\frac{\Gamma(\frac{1}{2}-\epsilon)}{\pi^{\frac{3}{2}-\epsilon}}L^{{2}\epsilon}
\eea
and
\bea
\mathcal{I}_2
&=&-\frac{\Gamma(\frac{3}{2}-\epsilon)\cos\phi}{\pi^{\frac{3}{2}-\epsilon}}\int_0^Ld\tau_1\int_{-L}^0d\tau_2\frac{\tau_1-\tau_2}{(\tau_1^2+\tau_2^2-2\tau_1\tau_2\cos(2\phi))^{3/2-\epsilon}}\nn
&=&-\frac{\Gamma(\frac{1}{2}-\epsilon)}{\pi^{\frac{3}{2}-\epsilon}}L^{2\epsilon}\left[\frac{\sec\phi}{4\epsilon}-\frac12\sec\phi\log(1+\sec \phi)+{\mathcal O}(\epsilon)\right],\eea
Here  we have 
 used the identities
\bea
&&u_{+ R(L)}\gamma_\mu u_{-R(L)} x_{12}^\mu=\ii\tau_{12},\quad u_{+ R(L)}\gamma_\mu u_{-L(R)} x_{12}^\mu=\ii\tau_{12}\cos(\phi).
\eea

\begin{figure}[htbp]
  \centering
  \subfigure[]{\includegraphics[height=0.16\textwidth]{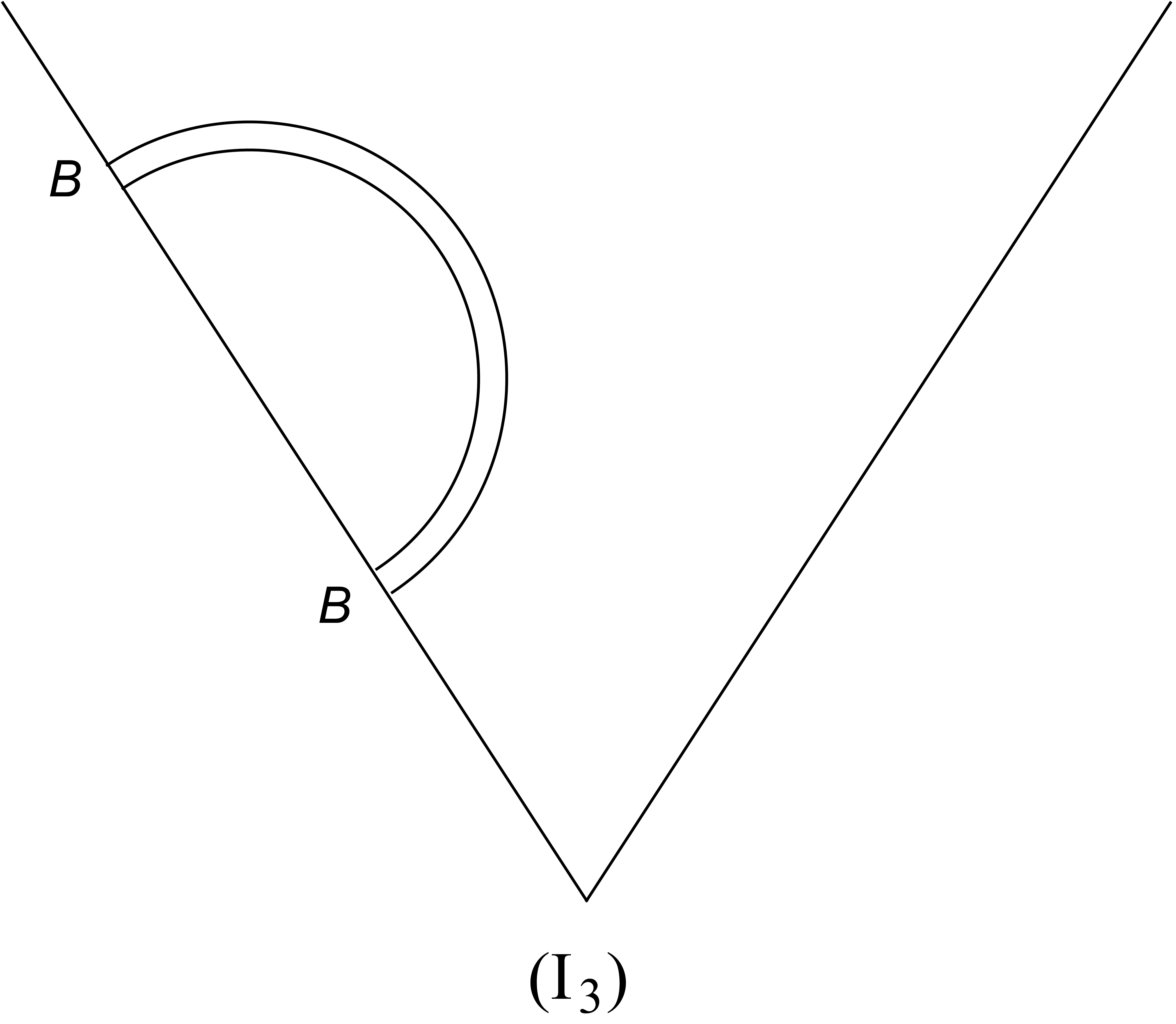}\label{I3}}~~
  \subfigure[]{\includegraphics[height=0.16\textwidth]{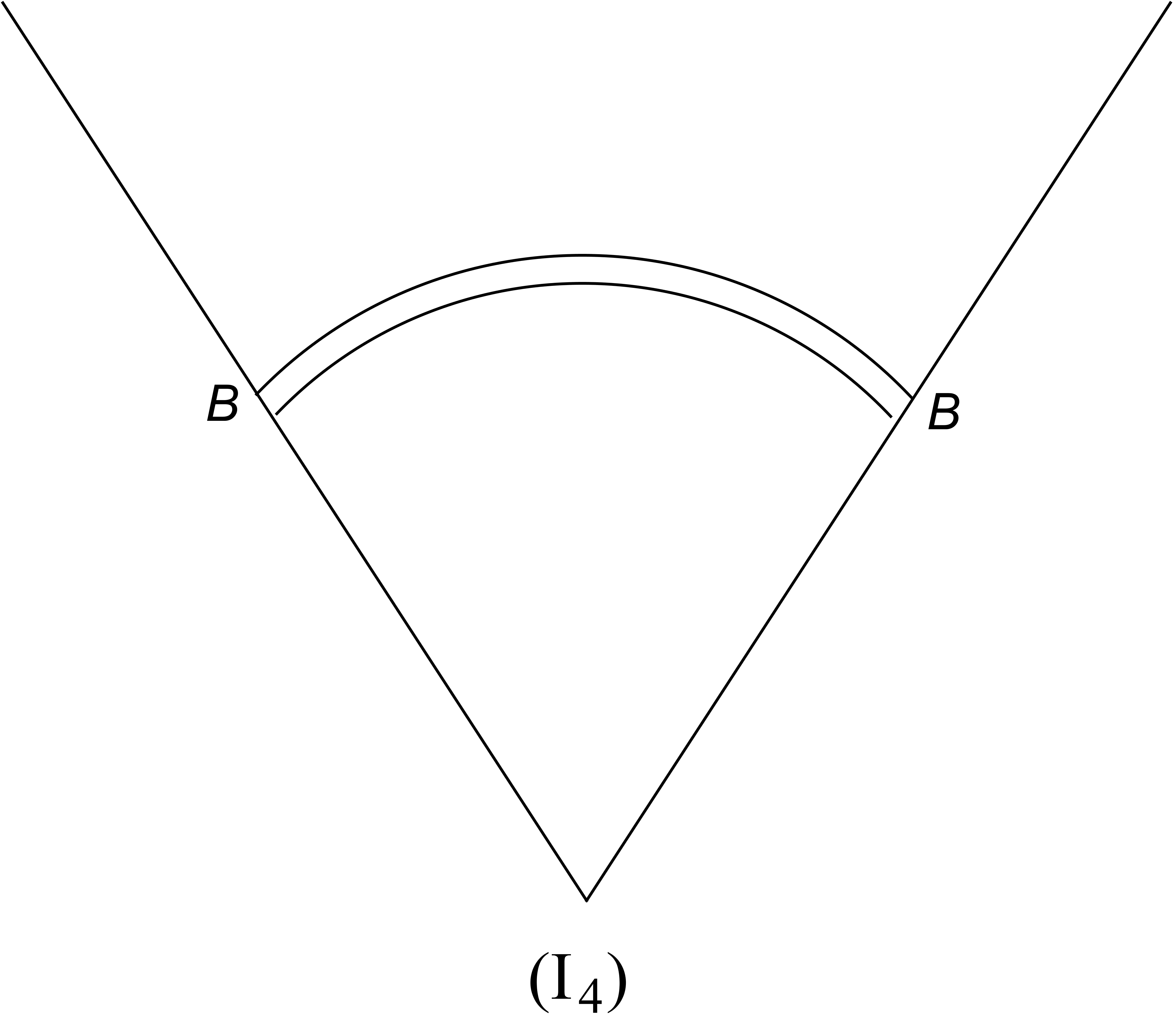}\label{I4}}~~
  \subfigure[]{\includegraphics[height=0.16\textwidth]{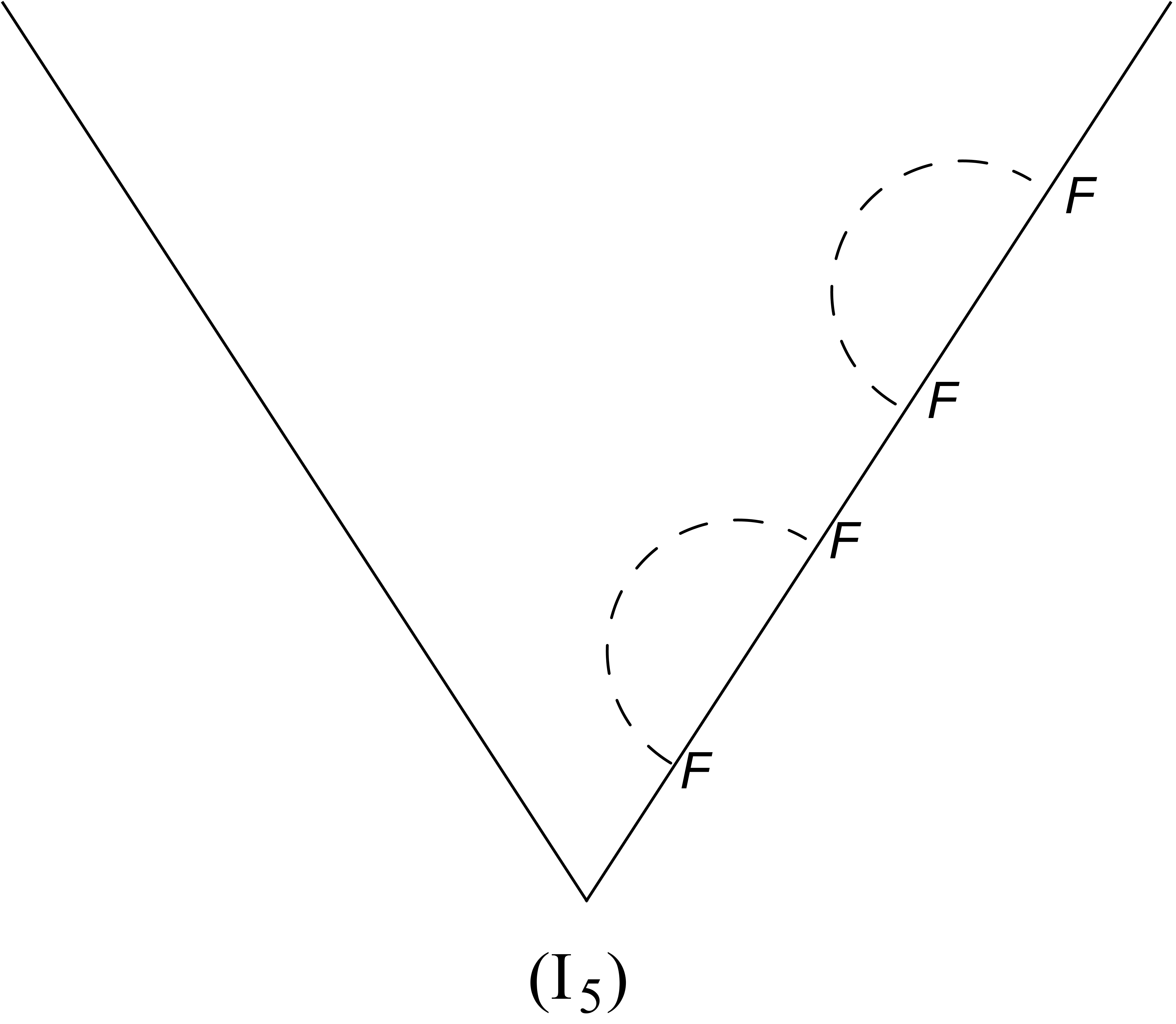}\label{I5}}
   \subfigure[]{\includegraphics[height=0.16\textwidth]{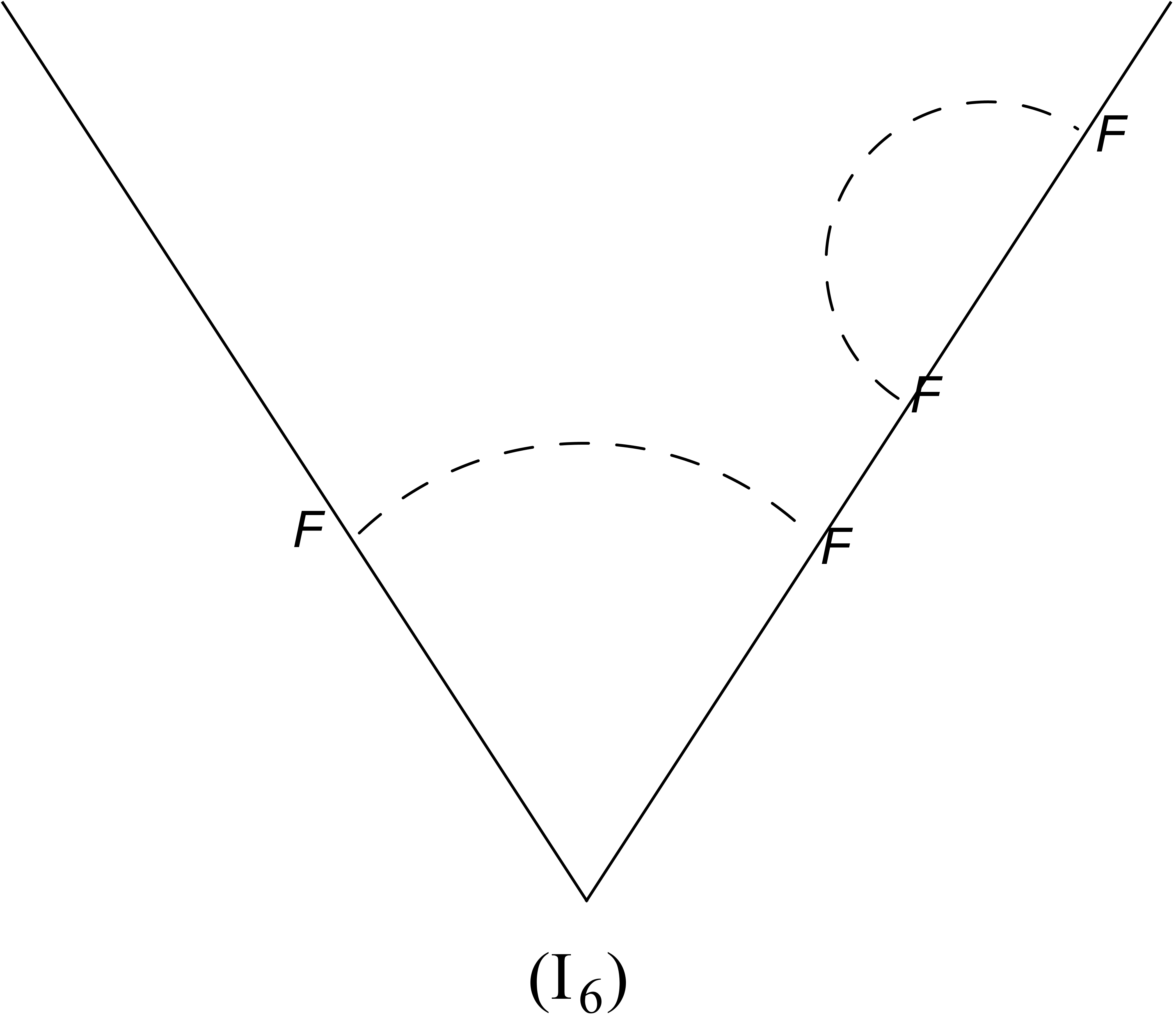}\label{I6}}~~ \\
   \subfigure[]{\includegraphics[height=0.16\textwidth]{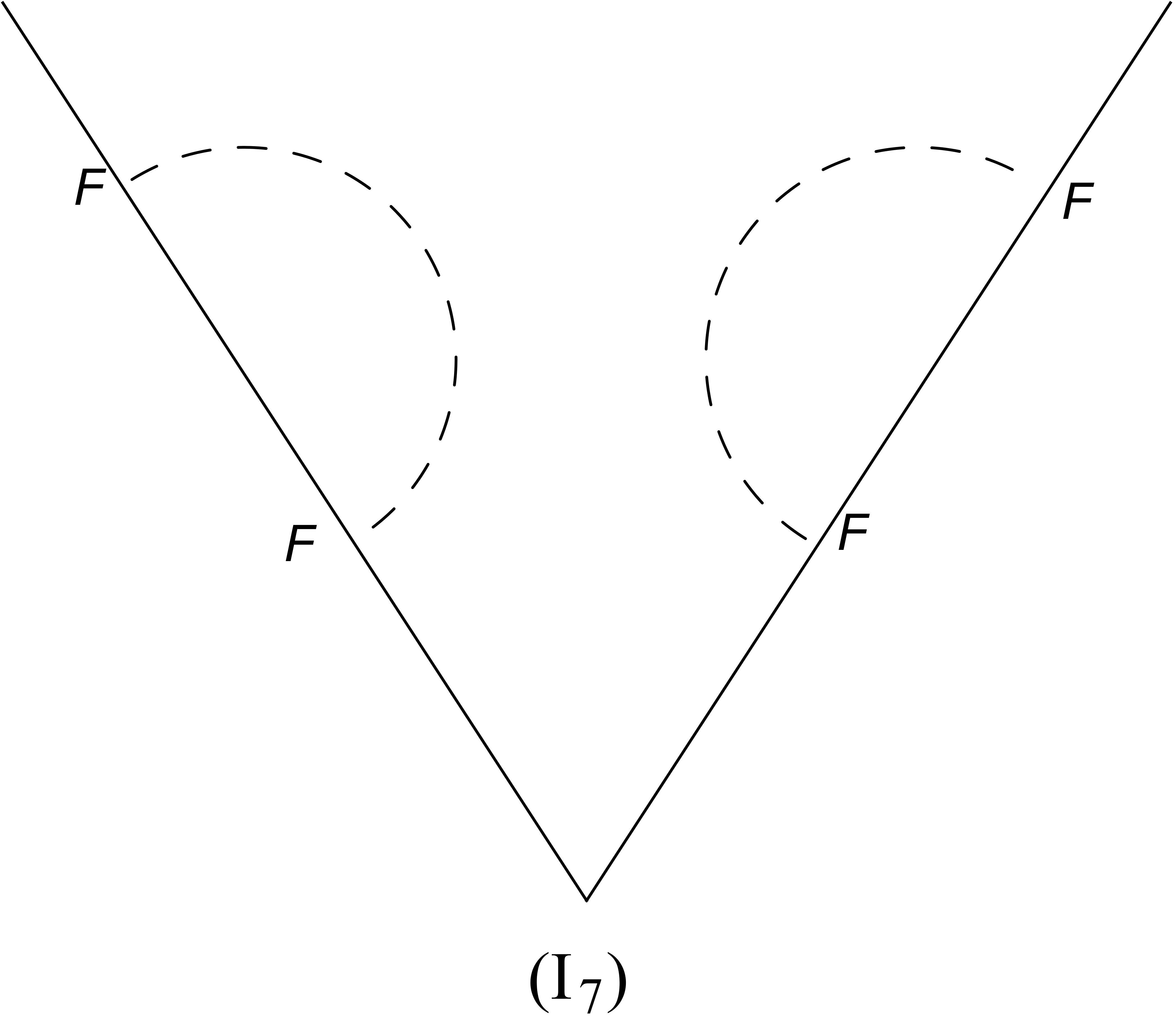}\label{I7}}~~
  \subfigure[]{\includegraphics[height=0.16\textwidth]{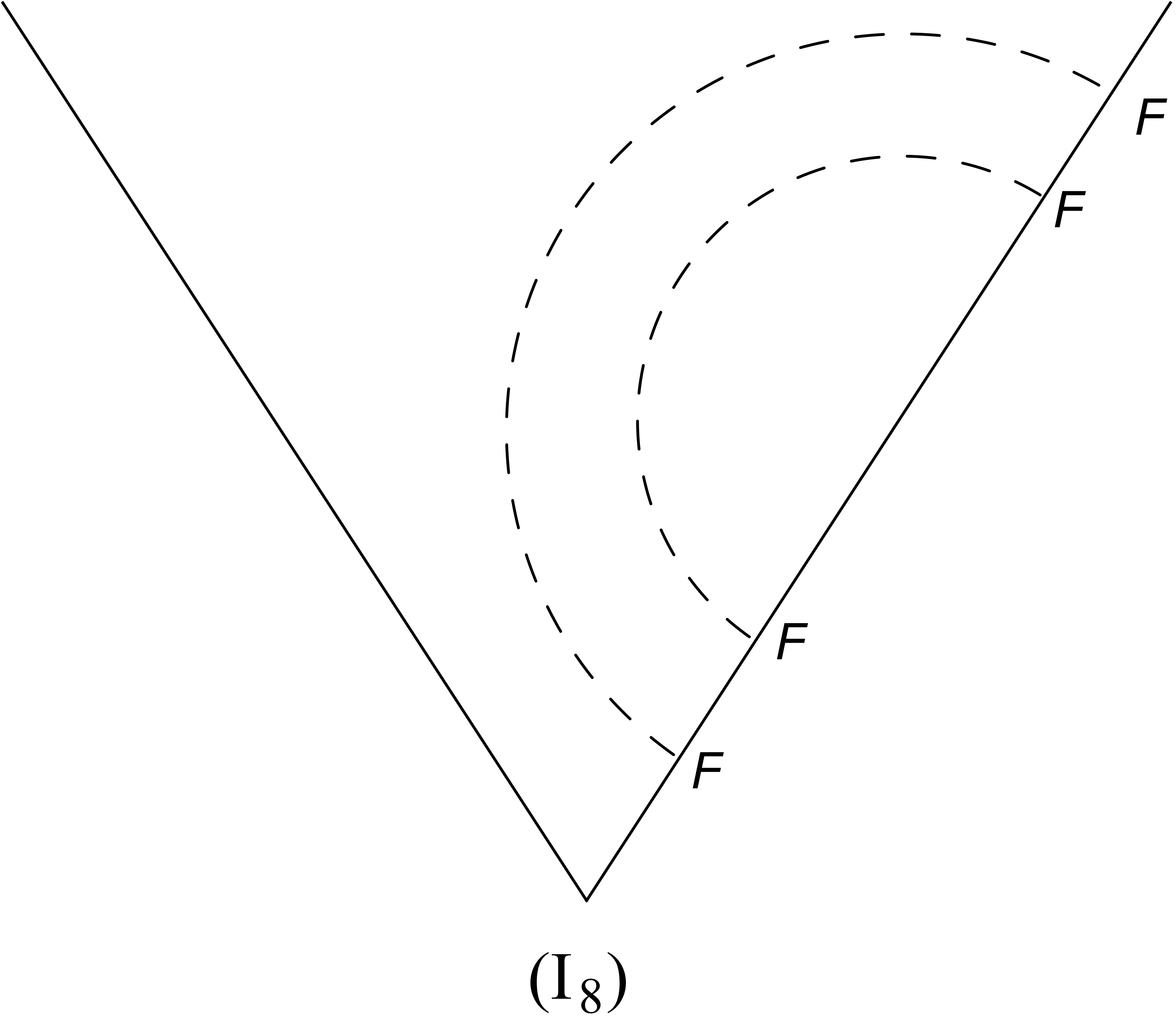}\label{I8}}~~
  \subfigure[]{\includegraphics[height=0.16\textwidth]{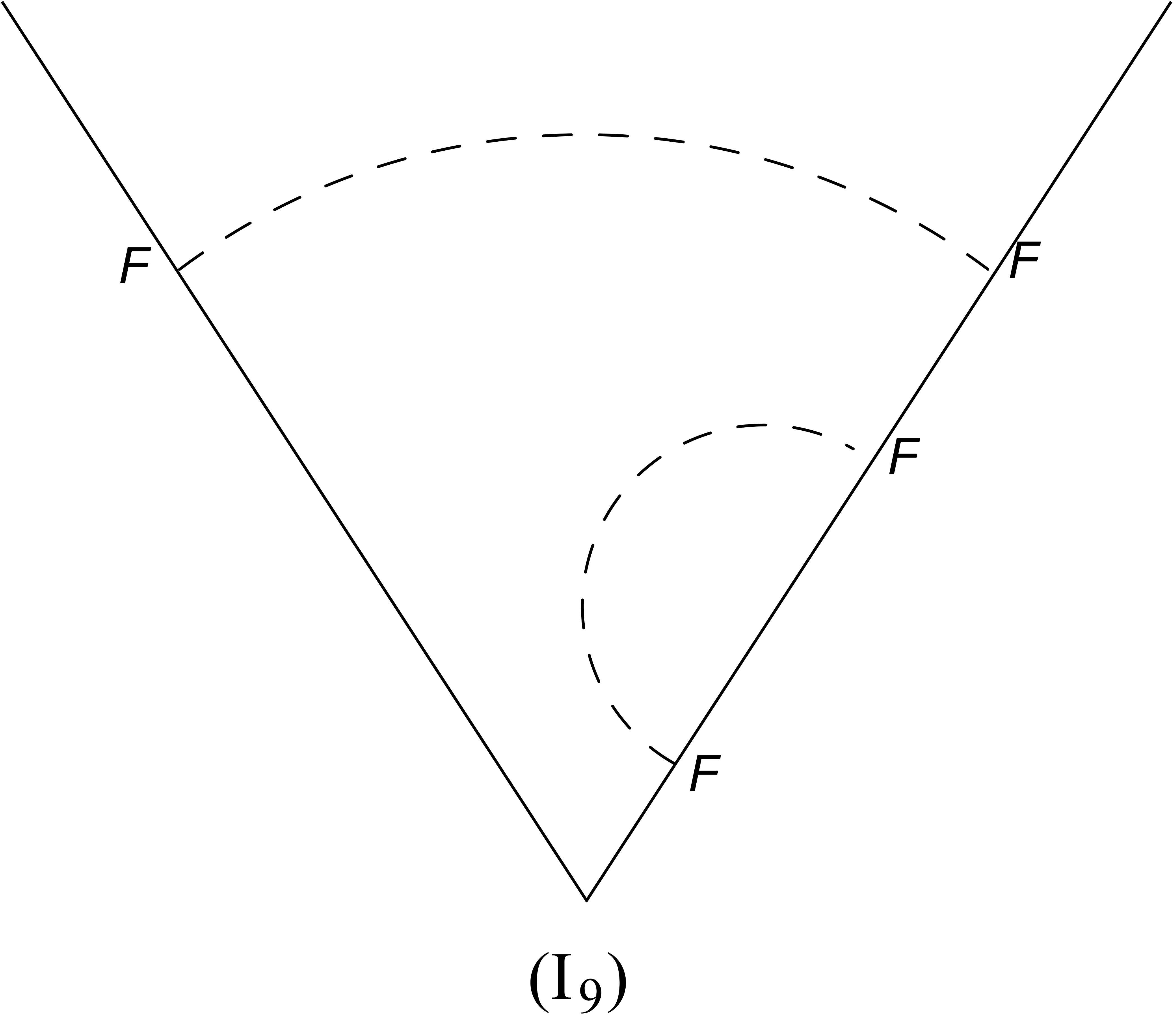}\label{I9}}
   \subfigure[]{\includegraphics[height=0.16\textwidth]{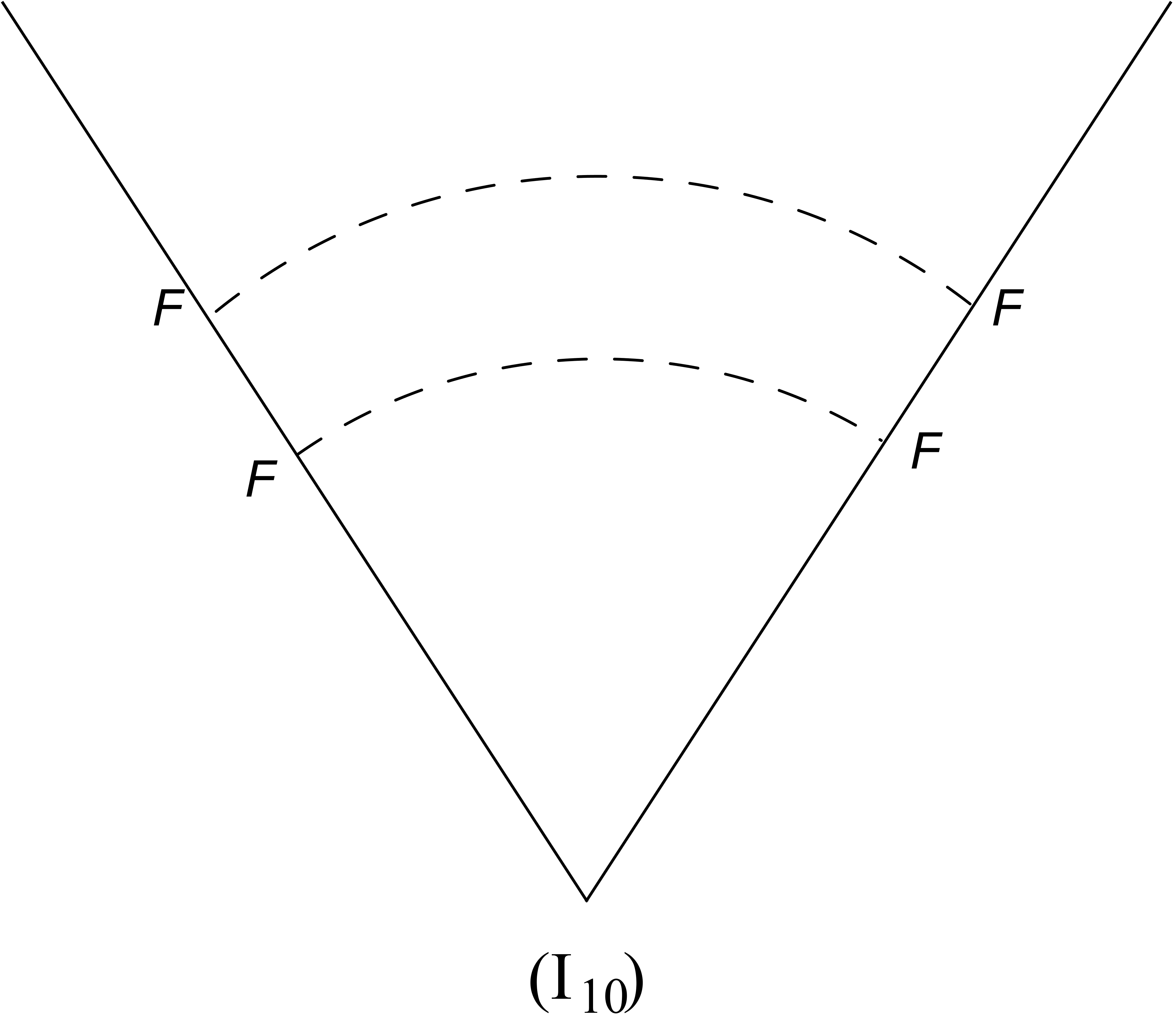}\label{I10}}~~
  \caption{Two-loop Feynman diagrams for cusp.}
\end{figure}
The two-loop contributions are from the diagrams with two scalar propagators (fig.~\ref{I3}, fig.~\ref{I4}) and two fermion propagators (figs.~\ref{I5}-\ref{I10}). The scalar contributions were not calculated in \cite{GenCusp} separately and were combined with the contributions from the one-loop corrected vector propagators. The scalar contributions are given by
\bea
W_{2B}
&=&\sum_{a, b, c}\left(\bar{m}^{ab}_in^i_{ba}\bar{m}^{ac}_jn^j_{ca}+\bar{m}^{ba}_in^i_{ab}\bar{m}^{ca}_jn^j_{ac}\right)N_aN_bN_c ( 2 ~{\mathcal I}_{3}+\mathcal{I}_4),
\eea
with
\bea
\mathcal{I}_3
&=&\frac{\Gamma^2(\frac{1}{2}-\epsilon)}{16 \pi^{3-2\epsilon}}\int_{-L<\tau_2<\tau_1<0}d\tau_1d\tau_2 \frac{1}{(\tau_1-\tau_2)^{2-4\epsilon}}\nn
&=&-\frac{\Gamma^2(\frac{1}{2}-\epsilon)}{16 \pi^{3-2\epsilon}}L^{4\epsilon}\left(\frac1{4\epsilon}+\mathcal{O}(1)\right) 
\eea
and
\bea
\mathcal{I}_4
&=&\frac{\Gamma^2(\frac{1}{2}-\epsilon)}{16 \pi^{3-2\epsilon}}\int_0^L d\tau_{1}\int_{-L}^0d\tau_2\frac{1}{(\tau_1^2+\tau_2^2-2\tau_1\tau_2\cos2\phi)^{1-2\epsilon}}\nn
&=&\frac{\Gamma^2(\frac{1}{2}-\epsilon)}{16 \pi^{3-2\epsilon}}L^{4\epsilon}\int_0^1 dx \int_0^1 dy\frac{1}{(x^2+y^2+2xya)^{1-2\epsilon}} \nn
&=&\frac{\Gamma^2(\frac{1}{2}-\epsilon)}{16 \pi^{3-2\epsilon}}L^{4\epsilon} \left(\frac{\phi}{2\epsilon \sin2\phi}+\mathcal{O}(1)\right).
\eea
Here we have defined a shorthand variable $a\equiv \cos(2\phi)$.
For details we evaluate the last integral in the following way,
\bea \label{trick}
&&\int_0^1 dx \int_0^1 dy\frac{1}{(x^2+y^2+2xya)^{1-2\epsilon}}=2\int_0^ydx \int_0^1 dy (x^2+y^2+2a x y)^{2\epsilon-1}\nn
&&=2\int_0^y\frac{dx}{y}\int_0^1dy y^{4\epsilon-1}((x/y)^2+1+2a x/y)^{2\epsilon-1}\nn
&&=2 \int_0^1 d\alpha(\alpha^2+1+2a \alpha)^{2\epsilon-1}\int_0^1 dyy^{4\epsilon-1}=\frac{1}{2 \epsilon}\frac{\phi}{\sin2\phi}+\mathcal{O}(1).
\eea

\noindent Lastly  we analyze the fermionic contributions which are 
\bea
&&W_{2F}={\color{red}}\sum_{a, b, c}\left(\bar{m}^{ab}_in^i_{ba}\bar{m}^{ac}_jn^j_{ca}+\bar{m}^{ba}_in^i_{ab}\bar{m}^{ca}_jn^j_{ac}\right)N_aN_bN_c(I_s+I_t),\nn
&&I_s=2\mathcal{I}_5+2\mathcal{I}_6+\mathcal{I}_7,\nn
&&I_t=2\mathcal{I}_8+2\mathcal{I}_9+\mathcal{I}_{10},
\eea
with

\bea
\mathcal{I}_5&=&\frac{\Gamma^2(\frac{3}{2}-\epsilon)}{4\pi^{3-2\epsilon}}\int_{0<\tau_4<\cdots<\tau_1<L}  \frac{\prod_{i=1}^4d\tau_i}{(\tau_1-\tau_2)^{2-2\epsilon}(\tau_3-\tau_4)^{2-2\epsilon}}\nn
&=&\frac{\Gamma\left(\frac12-\epsilon\right)^2}{16\pi^{3-2\epsilon}}L^{4\epsilon}(\frac1{4\epsilon^2}+\mathcal{O}(1)).\eea

\bigskip

\bea
\mathcal{I}_6&=&\frac{\Gamma^2(\frac{3}{2}-\epsilon)}{4\pi^{3-2\epsilon}} \cos\phi
\int_{-L}^0d\tau_4 \int_{0<\tau_3<\tau_2<\tau_1<L} \prod_{i=1}^3 d\tau_i \frac{1}{(\tau_1-\tau_2)^{2-2\epsilon}} \nn
&\times&\frac{\tau_3-\tau_4}{({\tau_3^2+\tau_4^2-2\tau_3\tau_4 \cos 2\phi})^{3/2-\epsilon}}\nn
&=&\frac{\Gamma^2(\frac{1}{2}-\epsilon)}{32\pi^{3-2\epsilon}}L^{4\epsilon}\left[-\frac{\sec\phi}{2\epsilon^2}+\frac{\sec\phi\log(1+\sec\phi)}{\epsilon}+\mathcal{O}(1)\right].
\eea
\bea
\mathcal{I}_7&=&\mathcal{I}_1^2/4=\frac{\Gamma\left(\frac12-\epsilon\right)^2}{64\pi^{3-2\epsilon}\epsilon^2}.
\eea

\bigskip

\bea
\mathcal{I}_8&=&\frac{\Gamma^2(\frac{3}{2}-\epsilon)}{4\pi^{3-2\epsilon}}\int_{0<\tau_4\cdots<\tau_1<L}\frac{\prod_{i=1}^4d\tau_i}{((\tau_1-\tau_4)(\tau_2-\tau_3))^{2-2\epsilon}}\nn
&=&\frac{\Gamma^2(\frac{3}{2}-\epsilon)}{16\pi^{3-2\epsilon}}L^{4\epsilon}\left(\frac1{8\epsilon^2}+\frac1{4\epsilon}+\mathcal{O}(1)\right).
\eea

\bigskip

\bea
 \mathcal{I}_9&=&\frac{\Gamma^2(\frac{3}{2}-\epsilon)}{4\pi^{3-2\epsilon}} \cos\phi
 \int_{-L}^0d\tau_4 \int_{0<\tau_3<\tau_2<\tau_1<L} \prod_{i=1}^3 d\tau_i \frac{1}{(\tau_2-\tau_3)^{2-2\epsilon}} \nn
 &\times&\frac{\tau_1-\tau_4}{({\tau_1^2+\tau_4^2-2\tau_1\tau_4 \cos 2\phi})^{3/2-\epsilon}}\nn
  &=&\frac{\Gamma^2(\frac{1}{2}-\epsilon)}{32\pi^{3-2\epsilon}}L^{4\epsilon} \left[-\frac{\sec\phi}{4\epsilon^2}+\frac1\epsilon\sec\phi\log(1+\sec\phi)+\frac1{2\epsilon}\sec\phi\log(\cos\phi)+\mathcal{O}(1)\right].
\eea

\bigskip

\bea
\mathcal{I}_{10}&=&\frac{\Gamma^2(\frac{3}{2}-\epsilon)}{4\pi^{3-2\epsilon}}\frac{\cos^2\phi}{(1+\cos2\phi)^2}\frac{L^{4\epsilon}}{(2\epsilon-1)^2}  \int_{-L<\tau_4<\tau_3<0}d\tau_3d\tau_4\int_{0<\tau_2<\tau_1<L}d\tau_1d\tau_2\nn
&& s(\tau_1,\tau_4)s(\tau_2,\tau_3)\nn
&=&\frac{\Gamma^2(\frac{1}{2}-\epsilon)}{32\pi^{3-2\epsilon}}\sec^2(\phi)L^{4\epsilon}\left[\frac{1}{{4}\epsilon^2}-\frac{1}{\epsilon}\log[1+\sec\phi]
-{\frac{1}{\epsilon}\frac{\phi\cos^2\phi}{\sin2\phi}}+\mathcal{O}(1)\right],
\eea
where we have defined $s(x,y)=(\p_x-\p_y)(x^2+y^2-2xy\cos(\phi))^{\epsilon-1/2}$.

\bigskip

Summing up all the contributions, we end up with the result up to two-loop level as\footnote{We have replaced $L$ with $\mu L$, where $\mu$ is some mass scale to make the quantity $\mu L$ dimensionless.}
\bea\label{CuspFinal}
&&\langle W\rangle=N_1+N_2+(\mu L)^{2\epsilon}N_1N_2\sum_{a\ne b}(\bar{m}_i^{ab}n_{ba}^i)\frac{\Gamma(\frac{1}{2}-\epsilon)}{\pi^{\frac{3}{2}-\epsilon}}\left(\frac{1}{2\epsilon}-\frac{\sec\phi}{4\epsilon}
\right.\nn
&&\quad \left.+\frac{\sec\phi}{2}\log(1+\sec\phi)\right)+(\mu L)^{4\epsilon}N_1 N_2(N_1+N_2)\sum_{a\ne b}(\bar{m}_i^{ab}n_{ba}^i)^2\Big\{\frac{\Gamma^2(\frac{1}{2}-\epsilon)}{16\pi^{3-2\epsilon}}\times\nn
&&\qquad\times \left[\frac{1}{\epsilon^2}(1-\frac{3\sec\phi}{4}
+\frac{\sec^2\phi}{8})+\frac1\epsilon \sec\phi\left(2\log(1+\sec\phi)+\frac12\log\cos\phi\right.\right.\nn
&&\left.\left.\qquad-\frac12\sec\phi \log(1+\sec\phi)+\mathcal{O}(1)\right)\right]\Big\}.
\eea

Notice that the contributions from exchanging two scalars are  cancelled by a part of  the contributions from exchanging two fermions. As the case of the ABJM theory, there exists a double pole in the expression (\ref{CuspFinal}) which is from the exchanges of the fermions. For the case at hand, there are only such diagrams which contribute to the double pole. From this result we can extract a generalized potential $V_N$. Recall that the BPS constraints in eq.~(\ref{BPSconditionNeq2}) admit the solutions with $\bar{m}_i^{21}=n^i_{12}=0$ or $\bar{m}_i^{12}=n^i_{21}=0$ which lead to nontrivial cusp operators. Without loss of generality, we consider the first case. From the definition of $V_N$ \cite{GenCusp},
\be \langle W\rangle= N_1\exp(V_{N_2})+N_2\exp(V_{N_1}).\ee
we can get the divergent part of $V_N$  as
\bea V_N&&=N \bar{m}_i^{12}n^i_{21} (\mu L)^{2\epsilon} \frac{\Gamma(\frac12-\epsilon)}{4\pi^{3/2-\epsilon}}\left(\frac1\epsilon-\frac{\sec\phi}{2\epsilon}{+\sec\phi\log(1+\sec\phi)}\right)\nn
&&+
N^2(\bar{m}_i^{12}n_{21}^i)^2\frac{\Gamma(\frac12-\epsilon)^2}{16\pi^{3-2\epsilon}}(\mu L)^{4\epsilon}\left(\frac{1}{2\epsilon^2}-\frac{\sec\phi}{4\epsilon^2}+
\frac{1}{\epsilon}\sec\phi\log(1+\sec\phi)\right.\nn
&&\left.+\frac{1}{2\epsilon}\sec\phi\log\cos\phi\right). \eea
Notice that there are $1/\epsilon^2$ terms in $(\mu L)^{4\epsilon}$ even in the straight line limit. This is different from the ABJM case and is related to the fact that no diagrams with vertices appears at two loops in the fishnet theory. It indicates that there may be a better way to extract $V_N$
from $\langle W\rangle$.\footnote{It was found  in \cite{1603.00541} that the prescription in \cite{GenCusp} fails starting at three-loop order. An alternative prescription which works better at higher-loop order in the ladder limit was provided in \cite{1603.00541}. This prescription is identical to the one in \cite{GenCusp} up to two-loop order. We would like to thank
Michelangelo Preti for discussions on this point.} We leave this problem for further studies.

 As discussed  in \cite{GenCusp}, the suitable renormalization condition  is that when $\phi=0$ the renormalized $V_N$ should vanish\footnote{Recall that we take the same $\bar{m}_i^{ab}$'s and $n^i_{ab}$ for the left and right part of the cusp.}.
From this condition,  we get the renormalized generalized potential
\bea V_N^{ren}&=&V_N-V_N|_{\phi=0}\nn
&&=N \bar{m}_i^{12}n^i_{21} (\mu L)^{2\epsilon} \frac{\Gamma(\frac12-\epsilon)}{4\pi^{3/2-\epsilon}}\left(\frac{\sec\phi}{2\epsilon}-\frac1{2\epsilon}{-\sec(\phi)\log(1+\sec(\phi))}+{\log2} \right)\nn
&&+
N^2(\bar{m}_i^{12}n_{21}^i)^2\frac{\Gamma(\frac12-\epsilon)^2}{16\pi^{3-2\epsilon}}(\mu L)^{4\epsilon}\left(-\frac{\sec\phi}{4\epsilon^2}+\frac{1}{4\epsilon^2}+
\frac{{1}}{\epsilon}\sec\phi\log(1+\sec\phi)\right.\nn
&&\left.-{\frac 1\epsilon \log 2}+\frac{1}{2\epsilon}\sec\phi\log\cos\phi\right).\eea
The so-called universal cusp anomalous dimension,   $\gamma_{cusp}$, is obtained from the large imaginary $\phi$ limit \cite{IR}:
 \be\gamma_{cusp}=-\lim_{\phi\to\infty} \frac{2\epsilon (V_N^{ren}|_{\phi\to i\phi})}{\phi}.\ee
If we still use  prescription  despite of the existence of $1/\epsilon^2$ terms,
we get $\gamma_{cusp}=0$ at two-loop order in our fishnet theory. 

\section{Conclusion and Discussions}

In this work, we studied the  loop operators in the three-dimensional $\mathcal{N}=2$ fishnet theories. This kind of theories was obtained by taking appropriate double scaling limit of the $\gamma$-deformed ABJM theory.
The action of this 3D ${\mathcal N}=2$ fishnet theory includes the kinetic terms and potential terms of the scalars and fermions, with the gauge sector got decoupled. It is not the simplest fishnet theory, but the presence of supersymmetries makes it attractive in studying the BPS  loop operators
in a setup simpler than the Chern-Simons-matter theories. We constructed the BPS line and loop operators, and  computed the VEV of circular BPS operators perturbatively. We got some numerical evidence to suggest that the one-loop and two-loop contributions are vanishing. This is in  consistency with the prediction from the localization, under the assumption that our loop operators are truly BPS at the quantum level.

During the perturbative computations we proposed a new regularization scheme at framing $-1$ which takes good care of the spinors in the definition of the fermionic BPS loop operators. The new scheme fits better into the spirit that the point-splitting should be consistent with the Hopf fibration used in the localization. However this new scheme makes the already complicated  computations at framing $-1$ more messy. It would be nice to further develop effective tools for such kind of computations in this scheme  numerically or even analytically. We hope this new scheme will be helpful in doing perturbative calculations in the usual supersymmetric Chern-Simons-matter theories. It could be useful to probe whether these classically BPS loop operators and other similar Wilson loops in super-Chern-Simons-matter with less supersymmetries are truly supersymmetric at the quantum level. 

Furthermore, we discussed the line operators with a cusp and studied their generalized cusp anomalous dimensions. We found that the two-loop generalized potential defined as in \cite{GenCusp} has a double pole in the expansion of $1/\epsilon$. It will be interesting to see if there are other prescription for the generalized potential  which leads  to the results with  only a simple pole. Put this issue aside, we found that
the universal  cusp anomalous dimension vanishes.

There has been little progress in the study of open chain from the (cusped) Wilson lines in the ABJM theory. It would
be valuable to study this hard problem in the simple setup here. That means to insert composite operator in the (cusped) line operators in
the 3D ${\mathcal N}=2$ fishnet theory and then investigate whether the open spin chain obtained here is integrable or not. We leave this problem
for future study.

\section*{Acknowledgment}

We would like to thank Jue Hou for his participation in the early stage of this project. We would also like to thank Hong Lu,  Gang Yang and Jiaju Zhang for very helpful discussions. The work of   JT  and BC was in part supported by NSFC Grant No.~11335012, No.~11325522 and No.~11735001.  The work of JW
was in part supported by NSFC Grant No.~11975164, No.~11935009 and  No.~11575202.  The diagrams were drawn using JaxoDraw \cite{Binosi:2003yf} which was based on
Axodraw \cite{Vermaseren:1994je}.

\appendix

\section{The Lagrangian of ABJM theory and its $\gamma$-deformation}
\label{ABJMaction}
\renewcommand{\theequation}{A.\arabic{equation}}
\setcounter{equation}{0}
In our convention the Lagrangian of ABJM theory  with gauge group $U(N)\times U(N)$ reads
\bea
&&\mathcal{L}_{ABJM}=\mathcal{L}_{CS}+\mathcal{L}_k+\mathcal{L}_p+\mathcal{L}_{Y},\nn
&&\mathcal{L}_{CS}=\frac{k}{4\pi}\epsilon^{\mu\nu\rho}\text{Tr}\Big(A_\mu \p_\nu A_\rho+\frac{2i}{3}A_\mu A_\nu A_\rho-B_\mu \p_\nu B_\rho-\frac{2i}{3}B_\mu B_\nu B_\rho\Big),\nonumber\\
&&\mathcal{L}_k=\text{Tr}(-D_\mu \bar{\phi}^I D^\mu \phi_I +i\b{\psi}_I \gamma^\mu D_\mu \psi^I),\nonumber\\
&&\mathcal{L}_p=\frac{4\pi^2}{3k^2}\text{Tr}(\phi_I \b{\phi}^I \phi_J \b{\phi}^J \phi_K \b{\phi}^K+\phi_I\b{\phi}^J \phi_J \b{\phi}^K \phi_K \bphi^I+4 \phi_I \bphi^J \phi_K \bphi^I \phi_J \bphi^K\nn
&&-6\phi_I \bphi^J\phi_J\bphi^I\phi_K\bphi^K),\nn
&&\mathcal{L}_{Y}=\frac{2\pi \ii}{k}\text{Tr}(\phi_I \bphi^I \psi^J\bpsi_J-2\phi_I\bphi^J\psi^I\bpsi_J-\bphi^I\phi_I\bpsi_J\psi^J+2\bphi^I \phi_J\bpsi_I \psi^J\nn
&&\quad\qquad +\epsilon^{IJKL}\phi_I\bpsi_J\phi_K\bpsi_L-\epsilon_{IJKL}\bphi^I \psi^J\bphi^K \psi^L).
\eea
Here $A_\mu, B_\mu$ are the gauge fields corresponding to the first and the second $U(N)$, respectively. $\phi_I$ and $\psi^I$ are four scalars and four fermions in the  $(N, \bar{N})$  representation of the $U(N)\times U(N)$
and their covariant derivatives are given by
\bea
&&D_\mu \phi_I=\p_\mu \phi_I +\ii A_\mu \phi_I-\ii\phi_I B_\mu,\nn
&&D_\mu\bphi^I=\p_\mu \bphi^I-\ii\bphi^I A_\mu +\ii B_\mu \bphi^I,\nn
&&D_\mu \psi^I=\p_\mu \psi^I+\ii A_\mu \psi^I-\ii\psi^IB_\mu,\nn
&&D_\mu \bpsi_I=\p_\mu \bpsi_I-\ii \bpsi_I A_\mu+\ii B_\mu\bpsi_I.
\eea

After the $\gamma$--deformation, some of the interaction terms will acquire non-trivial phases. Among the bosonic potential terms, only the terms with $\phi_I \bphi^J \phi_K \bphi^I \phi_J \bphi^K\equiv(I,J,K)$ with $I, J, K$ different from each other  acquire non-trivial phases as follows,

\bea \label{3limit}
&&\sum_{\sigma\in S(123)} [e^{\ii\, \text{sign}(\sigma)(\gamma_1+\gamma_2+\gamma_3)}(\sigma(1), \sigma(2),\sigma(3))]\nonumber\\
&&+\sum_{\sigma\in S(124)}[e^{\ii\, \text{sign}(\sigma)(\gamma_1+\gamma_2-\gamma_3)}(\sigma(1), \sigma(2), \sigma(4))]\nonumber\\
&&+\sum_{\sigma\in S(234)}[e^{\ii\, \text{sign}(\sigma)(-\gamma_1+\gamma_2+\gamma_3)}(\sigma(2),\sigma(3),\sigma(4))]\nonumber\\
&&+\sum_{\sigma\in S(314)}[e^{\ii\, \text{sign}(\sigma)(\gamma_1-\gamma_2+\gamma_3)}(\sigma(3), \sigma(1), \sigma(4))],\eea
where $\text{sign}(\sigma)=1 \,(-1)$ for the even (odd) permutation $\sigma$.

Among the Yukawa-like interaction only the following obtain non-trivial phase factors,
\bea \label{2limit}
&&(I,J)\equiv \bphi^I \phi^J\bpsi_I\psi^J(\text{ or }\phi_I\bphi^J \psi^I \bpsi_J)\rightarrow \nn
&&e^{\frac{\ii}{2}(\gamma_1+\gamma_2)}(1, 2)+e^{-\frac{\ii}{2}(\gamma_1+\gamma_2)}(2, 1)+e^{\frac{\ii}{2}(\gamma_2+\gamma_3)}(2, 3)+e^{-\frac{\ii}{2}(\gamma_2+\gamma_3)}(3, 2)\nn
&&+e^{\frac{\ii}{2}(\gamma_1+\gamma_3)}(3, 1)+e^{-\frac{\ii}{2}(\gamma_1+\gamma_3)}(1, 3)+e^{\frac{\ii}{2}(\gamma_2-\gamma_3)}(4, 1)+e^{-\frac{\ii}{2}(\gamma_2-\gamma_3)}(1, 4)\nn
&&+e^{\frac{\ii}{2}(\gamma_3-\gamma_1)}(4, 2)+e^{-\frac{\ii}{2}(\gamma_3-\gamma_1)}(2, 4)+e^{\frac{\ii}{2}(\gamma_1-\gamma_2)}(4, 3)+e^{-\frac{\ii}{2}(\gamma_1-\gamma_2)}(3, 4),
\eea
and
\bea \label{1limit}
[I,J,K,L]\equiv && \epsilon_{IJKL}\bphi^I\psi^J \bphi^K \psi^L (\text{ or } \epsilon^{IJKL}\phi_I{\bpsi_J} \phi_K {\bpsi_L}) \rightarrow \nn  &&e^{\frac{\ii}{2}\gamma_1}([1, 3, 4, 2]+[4, 2, 1, 3]-[3, 4, 2, 1]-[2, 1, 3, 4])\nonumber \\
&&+e^{-\frac{\ii}{2}\gamma_1}(-[1, 2, 4, 3]-[4, 3, 1, 2]+[2, 4, 3, 1]+[3, 1, 2, 4])\nonumber \\
&&+e^{\frac{\ii}{2}\gamma_2}(-[1, 4, 3, 2]-[3, 2, 1, 4]+[2, 1, 4, 3]+[4, 3, 2, 1])\nonumber \\
&&+e^{-\frac{\ii}{2}\gamma_2}([1, 2, 3, 4]+[3, 4, 1, 2]-[2, 3, 4, 1]-[4, 1, 2, 3])\nonumber\\
&&+e^{\frac{\ii}{2}\gamma_3}(-[1, 3, 2, 4]-[2, 4, 1, 3]+[3, 2, 4, 1]+[4, 1, 3, 2])\nonumber \\
&&+e^{-\frac{\ii}{2}\gamma_3}([1, 4, 2, 3]+[2, 3, 1, 4]-[3, 1, 4, 2]-[4, 2, 3, 1]).
\eea
Notice  that terms like $(I, I)$ are unchanged.

\section{The propagators of the scalars and fermions }
\label{FeynmanRules}
\renewcommand{\theequation}{B.\arabic{equation}}
\setcounter{equation}{0}
In the Minkowski spacetime,  we have the following kinematic  terms  for the scalars and the spinors in the $\mathcal{N}=2$ notation \bea
&& \cL_k = \sum_{a,b}{\textrm Tr} ( - \partial_\m \bar Z_i^{(ba)} \partial^\m Z^i_{(ab)} + \ii \bar \z_i^{(ba)} \g^\m \partial_\m \z^i_{(ab)}  ).
\eea
By standard Wick rotation, these terms become
\bea\label{Lcomponents}
&& \cL_k = \sum_{a,b}{\textrm Tr}( \partial_\m \bar Z_i^{(ba)} \partial^\m Z^i_{(ab)} - \ii \bar \z_i^{(ba)} \g^\m \partial_\m \z^i_{(ab)}  ).
\eea
Then  the explicit expressions for the propagators at tree level in position space  are
\bea
&& \lag Z_{(ab)}^i{}_p{}^q(x) \bar Z^{(cd)}_j{}_r{}^s(y) \rag = \d_a^d\d_b^c\d^i_j\d_p^s \d_r^q
                                                                      \f{\G(\f12-\e)}{4\pi^{\f32-\e}}
                                                                      \f{1}{|x-y|^{1-2\e}}, \nn
&& \lag \z_{(ab)}^i{}_p{}^q{}_\a(x) \bar \z^{(cd)}_j{}_r{}^s{}^\beta(y) \rag = \ii \d_a^d\d_b^c\d^i_j\d_p^s \d_r^q
                                                                                       \f{\G(\f32-\e)}{2\pi^{\f32-\e}}
                                                                                       \f{\g_\m{}_\a{}^\beta(x-y)^\m}{|x-y|^{3-2\e}},
\eea
where the dimensional regularization with $d=3-2\epsilon$ has been used.

\end{document}